\documentstyle[prb,aps,eqsecnum,epsf,preprint,tighten]{revtex}

\begin{document}
\draft
\title{Fermi liquid theory: a renormalization group point of view  }
\author{ N. Dupuis\thanks{On leave from Laboratoire de Physique des Solides,
Universit\' e Paris-Sud,
91405 Orsay, France }}
\address{Department of Physics, University of Maryland, College Park, MD
20742-4111} 
\maketitle
\begin{abstract}
We show how Fermi liquid theory results can be systematically
recovered using a renormalization group (RG) approach. Considering a
two-dimensional system with a circular Fermi surface, we derive RG
equations at one-loop order for the two-particle vertex function $\Gamma $
in the limit of small momentum (${\bf Q}$) and energy ($\Omega $) transfer
and obtain  the equation which determines the collective modes of a Fermi
liquid. The density-density response function is also calculated. 
The Landau function (or, equivalently, the Landau parameters $F_l^s$ and
$F_l^a$) is determined by the fixed point value of the $\Omega $-limit of
the two-particle vertex function (${\Gamma ^\Omega }^*$). We show
how the results obtained at one-loop order can be extended to all orders in a
loop expansion. Calculating the quasi-particle life-time and renormalization
factor at two-loop order, we reproduce the results obtained from
two-dimensional bosonization or Ward Identities. We  
discuss the zero-temperature limit of the RG equations and the difference
between the Field Theory and the Kadanoff-Wilson formulations of the RG. We
point out the importance of $n$-body ($n\geq 3$) interactions in the latter.
\end{abstract}

\section{Introduction}
\label{sec:intro}

Since the original work of Landau,
\cite{Landau57,Landau59,Abrikosov63,Nozieres64}  the Fermi liquid 
theory (FLT) is one of the main basis of our understanding of interacting 
fermions.
The discovery of new materials showing strong deviations with respect to FLT, 
like high-$T_c$ superconductors, has revived interest in the microscopic
derivation of Landau's theory. In particular, this has motivated the
application of RG methods to interacting fermions in dimension $d\geq 2$.
\cite{Shankar91,Polchinski92,Shankar94,Chitov95,Feldman95,Dupuis95,Hewson94}

RG methods are well-known for one-dimensional interacting
fermions. \cite{Solyom79,Bourbon91} In these systems, the low-order
perturbation theory is characterized by two logarithmically singular and
interfering channels of correlation, namely, the particle-particle (Cooper)
and $2k_F$ particle-hole (Peierls) channels. This invalidates any RPA-like
approach which implicitly assumes the independence of the channels to
lowest order. The RG approach allows one to sum up the leading,
next-to-leading... logarithmic singularities in a consistent way. It has
been also successfully applied to quasi-one-dimensional
conductors (i.e. weakly coupled chains systems) where the interchain
coupling can lead either to a Fermi liquid behavior or to a state of broken
symmetry. \cite{Bourbon91,Bourbon88}
RG methods have also been used to study the instabilities of isotropic
two-dimensional systems with respect to superconductivity or charge (spin)
density wave. \cite{Schulz87} 
 While the independence of various channels of
correlation is in general expected to be a good approximation (therefore
allowing an RPA approach), for particular Fermi surfaces the situation can
become more complicated due to the presence of nesting, Van Hove
singularities... In such cases, the RG can be a useful tool to study the
instabilities of the system and to investigate in detail phenomena like
superconductivity induced by exchange of spin fluctuations.

In the above mentioned examples, one focuses on ``highly quantum'' degrees
of freedom corresponding to energies larger than the temperature. The 
perturbation theory is characterized in general by 
logarithmic singularities of the type $\ln (E_0/T)$ ($E_0$ being an
ultra-violet cut-off), which clearly shows that the temperature plays the
role of an infrared cut-off. This should be contrasted with the standard
diagrammatic derivation of FLT\cite{Landau59,Abrikosov63,Nozieres64} (which
in the following is
referred to as the microscopic FLT) where these ``quantum'' degrees of
freedom are in general not considered explicitly but simply included in the
definition of some regular low-energy effective interactions. FLT
concentrates on the Landau (or zero-sound (ZS)) channel (particle-hole pairs
at small total
momentum and energy) where the important degrees of freedom are known to be
within the thermal broadening of the Fermi surface. This latter property can
be seen from the polarization diagram (or particle-hole bubble) which is
proportional to $\partial n_F/\partial \epsilon $ to lowest order in
perturbation theory ($n_F(\epsilon )$ is the Fermi occupation factor). Thus
the role of temperature is to fix the typical energy scale rather than to
provide an infrared cut-off. This makes the application of RG methods in
that case somewhat different from the above mentioned cases.

The first attempt to recover (in detail) Landau's theory from a RG method is
due to Shankar. \cite{Shankar91,Shankar94} However, Shankar's discussion of FLT
rather relies on usual perturbative theory than on RG approach. Although 
RG arguments were used to identify the relevant couplings, the low-energy
effective degrees of freedom were indeed  explicitly integrated out by
means of standard diagrammatic calculations. Using a finite temperature
formalism, Chitov and S\'en\'echal have shown how FLT can be understood from
a RG approach. In particular, they have correctly analyzed (within the RG
framework) the singularities of the Landau channel which are at the heart of
the microscopic FLT. Chitov and
S\'en\'echal's analysis has recently been further developed and a detailed
connection between the microscopic FLT and the RG approach has
emerged. \cite{Dupuis95} 

The aim of this paper is to show how FLT results can be systematically
derived in a RG approach. On the one hand we derive in
detail the results reported in Ref.\ \onlinecite{Dupuis95}. On the other
hand, we present new results and discuss at length some particular points
concerning the application of RG methods to interacting fermions.  
In the next section, we recall some aspects of FLT. Our aim
is not to give an exhaustive summary of FLT, but to mention the main ideas
underlying the microscopic FLT while emphasizing some points which will turn
out to be crucial in the RG approach, such as the singularity of the
two-particle vertex function at small momentum and energy transfer or its
symmetry properties. In Sec.\ \ref{sec:1L}, we derive the RG equations at
one-loop order for the $Q$-limit ($\Gamma ^Q$) and $\Omega $-limit ($\Gamma
^\Omega $) of the forward scattering vertex. In order to respect the Fermi
statistics, the forward scattering zero-sound (ZS) and exchange (ZS$'$) are
both taken into account. As a result, we find that both the flows of $\Gamma
^Q$ and $\Gamma ^\Omega $ are non zero. We show that the antisymmetry of
$\Gamma ^Q$ under exchange of the two incoming or outgoing particles is
conserved under RG, while the antisymmetry of $\Gamma ^\Omega $ is lost. 
We solve (approximately) the RG equations to obtain a relation between the
fixed point values ${\Gamma ^Q}^*$ and ${\Gamma ^\Omega }^*$. We
then extend the RG equations to the case of finite momentum and
energy transfers and obtain the equation determining the
collective modes of a Fermi liquid. The standard results of FLT are
recovered if one identifies the Landau parameters $F_l^s$, $F_l^a$ with 
${\Gamma ^\Omega }^*$. We calculate the density-density correlation function
and discuss the zero temperature limit of the RG equations. In Sec.\
\ref{sec:afr}, we discuss in detail some subtle points concerning the
implementation of RG methods to interacting fermions. We point out the
importance of three-, four-, ...-body interactions in the Kadanoff-Wilson
(KW) formulation of the RG and discuss the differences between this approach
and the Field Theory (FT) approach. In Sec.\ \ref{sec:bol},
we show how the results obtained at one-loop order can be extended to all
orders. The quasi-particle life-time and renormalization factor are
explicitly calculated at two-loop order in Sec.\ \ref{sec:qpp}. The
scattering rate is found to be $\tau ^{-1}\sim T^2\ln T$, a result
previously known for a two-dimensional Fermi liquid. The expression for the
renormalization factor agrees with the one obtained from two-dimensional
bosonization or Ward Identities. We only consider in this paper
the case of a neutral system with short range interactions.  

\section{Some aspects of Fermi liquid theory}
\label{sec:FLT}

\subsection{Phenomenological approach }
\label{subsec:FLT-PA}

Landau's first approach to Fermi liquids is phenomenological. 
\cite{Landau57,Abrikosov63,Nozieres64} The main 
assumption is the existence of low-energy elementary excitations 
(quasi-particles) 
which can be put in a one-to-one correspondence with the elementary excitations
(``particles'' and ``holes'') of the non-interacting fermion gas. Landau 
further postulated that a weak perturbation applied to the system in its 
ground state induces a change of the total energy given by (from now on we 
consider spin one-half fermions)
\begin{equation}
\delta E=\sum _{{\bf K},\sigma } \epsilon ^0_{\bf K} \delta n_{{\bf K}\sigma }
+{1 \over {2\nu }} \sum _{{\bf K},{\bf K'},\sigma ,\sigma '}
f_{\sigma ,\sigma '}({\bf K},{\bf K'})
\delta n_{{\bf K}\sigma } \delta n_{{\bf K'}\sigma '} +O(\delta n ^3) \,,
\label{Elandau}
\end{equation}
where $\delta n_{{\bf K}\sigma }$ is the change in the occupation number of 
the quasi-particles of momentum ${\bf K}$ and spin $\sigma $, and $\nu $ the 
volume of the system. $\epsilon ^0_{\bf K}$ is the energy of a quasi-particle
in the absence of other excited quasi-particles. For states near the Fermi 
surface and in an isotropic liquid (the only case we shall consider in this
paper), it can be written as $\epsilon 
^0_{\bf K}=v_F(K-K_F)+\mu $ where 
$K_F$ is the Fermi wave-vector, $v_F$ the Fermi velocity of the 
quasi-particles and $\mu $ the chemical potential (we set $\hbar =k_B=1$ 
throughout the paper). The 
effective mass is defined by $m=K_F/v_F$. The second term of the rhs of 
(\ref{Elandau}) comes from the interaction between quasi-particles. 
For states very close to the Fermi surface, $K\simeq K_F$ and $K'\simeq K_F$, 
the Landau function $f_{\sigma ,\sigma '}$
is a function of the angle between ${\bf K}$ and ${\bf K'}$. If the spin 
dependent part of the quasi-particle interaction is due purely to exchange, 
$f_{\sigma ,\sigma '}$ can be written as
\begin{eqnarray}
f_{\sigma ,\sigma '}(\theta )&=& f^s(\theta )
+f^a(\theta ) \sigma \sigma '  
 \nonumber \\
&=& {1 \over {2N(0)}} \sum _{l=0}^\infty (F^s_l +F^a_l 
\sigma \sigma ') P_l(\cos \theta ) \,,
\label{Legendre}
\end{eqnarray}
where $N(0)=K_F^2/2\pi ^2v_F$ is the density of states per spin at the Fermi 
level and $\sigma =1$ $(-1)$ for up (down) spins. In the last line 
of (\ref{Legendre}), we have expanded $f^s(\theta )$ and $f^a(\theta )$ 
on the basis of Legendre polynomials $P_l(\cos \theta )$. Using 
(\ref{Elandau}) and a transport equation for the distribution function of the 
quasi-particles, one can relate the physical quantities to the Landau
parameters $F^s_l$ and $F^a_l$.\cite{masse} For example, 
one obtains for the specific heat
and the Pauli susceptibility, $C=C^0(1+{{F^s_1} \over 3})$ and $\chi _P= \chi 
_P^0/(1+F^a_0)$, where $C^0$ and $\chi ^0_P$ denote the corresponding
quantities  
in a free fermion gas. Eq.\ (\ref{Elandau}) holds only when the spin 
projection of the quasi-particles along a given axis is a good quantum number.
In a more general case, $\epsilon ^0$ and $\delta n$ should be considered as 
matrices in the spin variables. The Landau function then becomes a function 
$f_{\sigma _i}\equiv f_{\sigma _1\sigma _2,\sigma _3\sigma _4}$ of four spin 
variables. In the following, we shall rather use $f_{\sigma _i}$ than 
$f_{\sigma ,\sigma '}$. 

\subsection{Microscopic approach}
\label{subsec:FLT-MA}

\subsubsection{Bethe-Salpeter equation in the ZS channel}

\begin{figure}
\epsfysize8.5cm
\epsffile[98 167 514 624]{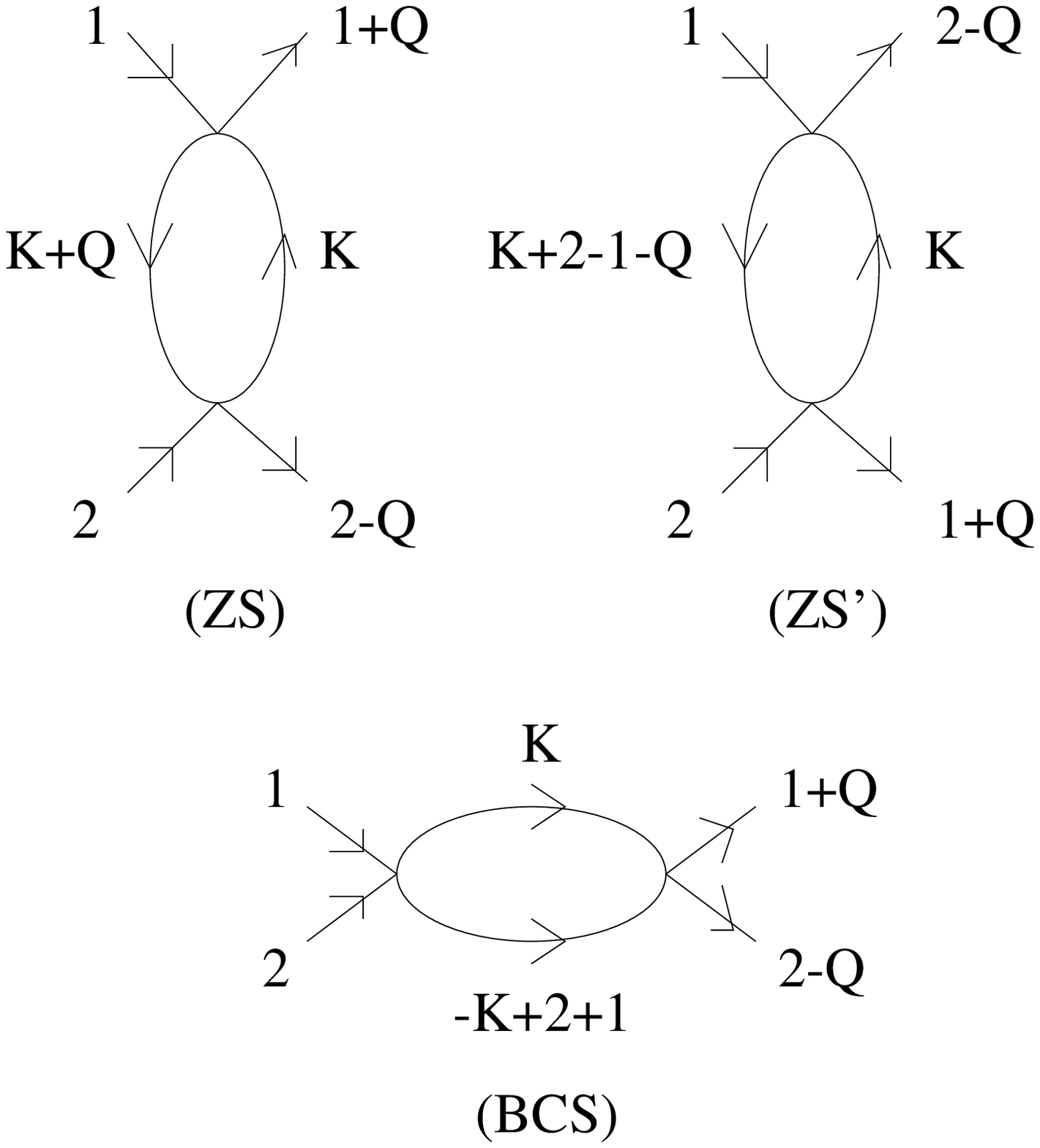}
\caption{Lowest order (one-loop) corrections to the two particle vertex 
function $\Gamma $. We use the simplified notation $1=\tilde K_1$,
$1+Q=\tilde K_1+\tilde Q$... }
\label{fig:corr_1loop}
\end{figure}

The foundations of FLT were rapidly established using field theoretical 
methods. \cite{Landau59,Abrikosov63,Nozieres64} 
The key microscopic quantity is the 
two-particle vertex function, which, together with the one-particle propagator 
$G$, plays an essential role in the theory of the Fermi liquid. We denote this 
quantity by $\Gamma _{\sigma _i}(\tilde K_1,
\tilde K_2;\tilde K_2-\tilde Q,\tilde K_1+\tilde Q)
\equiv \Gamma _{\sigma _i}(\tilde K_1,\tilde K_2;\tilde Q)$ with 
$\Gamma _{\sigma _i}\equiv \Gamma _{\sigma _1\sigma _2,\sigma _3\sigma _4}$. 
Here $\tilde K=({\bf K},\omega )$ where ${\bf K}$ is a three-component vector
and $\omega =\pi T(2n+1)$ ($n$ integer) a fermionic Matsubara frequency. 
${\bf Q}$ is the momentum transfer between the two particles and the bosonic
Matsubara frequency $\Omega =2\pi Tm$ ($m$ integer) 
the energy transfer. Landau noted that among the three lowest order
corrections for  
$\Gamma $  shown in Fig.\ \ref{fig:corr_1loop} (one-loop corrections), the
ZS graph plays a special role when $\tilde Q \rightarrow 0$ (and
$T\rightarrow 0$) since the product $G(\tilde K)G(\tilde K+\tilde Q)$
becomes singular in this limit if one assumes a quasi-particle form for the
one-particle propagator, i.e. $G(\tilde K)=z\lbrack i\omega
-v_F(K-K_F)\rbrack ^{-1}$ where $z$ is the quasi-particle renormalization
factor. (We do not consider the incoherent part $G_{\rm
inc}(\tilde K)$ which can easily be taken into account.) This motivated Landau
to organize the perturbation expansion of $\Gamma $ as follows. 
\cite{Landau59,Abrikosov63,Nozieres64} 
One first introduces the quantity $\tilde \Gamma $
defined as the sum of all diagrams which do not contain the singular product 
$G(\tilde K)G(\tilde K+\tilde Q)$. The exact two-particle vertex function is 
then determined by the following Bethe-Salpeter equation:
\begin{eqnarray}
\Gamma _{\sigma _i}
(\tilde K_1,\tilde K_2;\tilde Q)&=&
\tilde \Gamma _{\sigma _i}(\tilde K_1,\tilde K_2;\tilde Q) 
\nonumber \\
&& +{T \over \nu } \sum _{\sigma ,\sigma '} \sum _{\tilde K} \tilde 
\Gamma _{\sigma _1\sigma ',\sigma \sigma _4}(\tilde K_1,\tilde K;\tilde Q)
G(\tilde K)G(\tilde K+\tilde Q) 
\Gamma _{\sigma \sigma _2,\sigma _3\sigma '}
(\tilde K,\tilde K_2;\tilde Q) \,.
\label{Eqint}
\end{eqnarray}
Eq.\ (\ref{Eqint}) determines $\Gamma $ as a function of the
irreducible (with respect to the ZS channel) two-particle vertex function
$\tilde \Gamma $. In order to simplify it, we use the two following
properties: i) $\tilde \Gamma $ is a non-singular function of $\tilde Q$
(this is explicitly verified at one-loop order). This allows us to neglect
its $\tilde Q$-dependence at small $\tilde Q$. ii) The singularity of
$G(\tilde K)G(\tilde
K+\tilde Q)$ comes from $\tilde K$ in the vicinity of the Fermi surface. In
this area, the $K,\omega $-dependence of the $\Gamma $'s in (\ref{Eqint}) can
be neglected so that the energy and radial momentum integral can be done. 
\cite{Schulz95} Note that this amounts to decoupling the ZS channel from the
other channels. Using (i) and (ii), (\ref{Eqint}) transforms into
(for $T\to 0$)
\begin{eqnarray}
\Gamma _{\sigma _i}
(\tilde K_1,\tilde K_2;\tilde Q)&=&
\tilde \Gamma _{\sigma _i}(\tilde K_1,\tilde K_2) 
\nonumber \\
&& +z^2N(0) \sum _{\sigma ,\sigma '} \int {{d\Omega _{\bf \hat K}} 
\over {4\pi }} \tilde 
\Gamma _{\sigma _1\sigma ',\sigma \sigma _4}(\tilde K_1,\tilde K)
\Gamma _{\sigma \sigma _2,\sigma _3\sigma '}
(\tilde K,\tilde K_2;\tilde Q) {{v_F{\bf \hat K}\cdot {\bf Q}} 
\over {i\Omega -v_F{\bf \hat K} \cdot {\bf Q}}} \,, 
\label{Eqint1}
\end{eqnarray}
where ${\bf \hat K}={\bf K}/K$ is a unit vector and $d\Omega _{\bf \hat K}$
the corresponding angular integration. Eq.\ (\ref{Eqint1}) is the basis of the
microscopic derivation of FLT. Before solving it (Sec.\ \ref{subsubsec:LS}),
we discuss in the next section the importance of the ZS$'$ channel with
respect to the Fermi statistics.

\subsubsection{Symmetry considerations}

As pointed out by Mermin \cite{Mermin67,Lifshitz80}
(see also Ref.\ \onlinecite{Hewson94}), 
the ZS$'$ graph also becomes singular in the limit
${\bf K}_1-{\bf K}_2\rightarrow 0$ since it contains the product $G(\tilde K)
G(\tilde K+\tilde K_2-\tilde K_1-\tilde Q) \rightarrow  G(\tilde
K)G(\tilde K-\tilde Q)$. It is therefore necessary in this case
(at least in principle) 
to consider the ZS and ZS$'$ channels on the same footing and to add to the
rhs of (\ref{Eqint1}) the term
\begin{eqnarray}
& -{T \over \nu } \sum _{\sigma ,\sigma '} \sum _{\tilde K} &
\tilde \Gamma _{\sigma _1 \sigma ',\sigma \sigma _3} 
(\tilde K_1,\tilde K+\tilde K_2-\tilde K_1;\tilde K_2-\tilde K_1-\tilde Q)
G(\tilde K)G(\tilde K+\tilde K_2-\tilde K_1-\tilde Q)
\nonumber \\ && \times 
\Gamma _{\sigma \sigma _2,\sigma _4\sigma '}
(\tilde K,\tilde K_2;\tilde K_2-\tilde K_1-\tilde Q)
\nonumber \\
&\simeq &-z^2N(0) \sum _{\sigma ,\sigma '} \int {{d\Omega _{\bf \hat K}} 
\over {4\pi }} \tilde 
\Gamma _{\sigma _1\sigma ',\sigma \sigma _3}(\tilde K_1,\tilde K)
\Gamma _{\sigma \sigma _2,\sigma _4\sigma '}
(\tilde K,\tilde K_2;\tilde K_2-\tilde K_1-\tilde Q) 
\nonumber \\ && \times  
{{v_F{\bf \hat K}\cdot ({\bf K}_2-{\bf K}_1-{\bf Q})} 
\over {-i\Omega -v_F{\bf \hat K} \cdot ({\bf K}_2-{\bf K}_1-{\bf Q})}} \,, 
\label{Eqint2}
\end{eqnarray}
where the second line is obtained in the limit $Q\to 0$ and $\vert {\bf
K}_2-{\bf K}_1 \vert \ll T/v_F$ (and $T\to
0$). Note that the ZS graph alone does not respect the Fermi statistics. 
\cite{Mermin67,Lifshitz80,Hewson94} Indeed, if 
one exchanges the two incoming or outgoing lines, the ZS graph transforms into
the ZS$'$ graph and vice versa. The consideration of the ZS$'$ graph
when ${\bf K}_1-{\bf K}_2\rightarrow 0$ ensures that the 
antisymmetry properties of the vertex function are satisfied. From
(\ref{Eqint1},\ref{Eqint2}), we obtain 
\begin{equation}
\Gamma _{\sigma _1\sigma _1,\sigma _2\sigma _3}
(\tilde K_1,\tilde K_1;\tilde Q) =0 \,\,\, {\rm for}\,\,\, Q\neq 0 \,\,\, 
{\rm and}\,\,\, \Omega \neq 0 \,.
\label{Eqint3}
\end{equation}
Since the value of $\Gamma _{\sigma _i}(\tilde K_1,\tilde K_2;\tilde Q)$
for $\tilde Q\to 0$ depends in an essential way on the ratio $Q/\Omega $ (as
can be seen from 
(\ref{Eqint1},\ref{Eqint2})), we introduce the $Q$- and $\Omega $-limits of
the two-particle vertex function:
\begin{eqnarray}
\Gamma ^Q_{\sigma _i}(\tilde K_1,\tilde K_2) &=& \lim _{Q\rightarrow 0}
\Bigl \lbrack \Gamma _{\sigma _i}(\tilde K_1,\tilde K_2;\tilde Q) 
\Bigr \vert _{\Omega =0} \Bigr \rbrack \,, \nonumber \\
\Gamma ^\Omega_{\sigma _i}(\tilde K_1,\tilde K_2) &=& \lim _{\Omega 
\rightarrow 0} \Bigl \lbrack \Gamma _{\sigma _i}(\tilde K_1,\tilde K_2;
\tilde Q) \Bigr \vert _{Q=0} \Bigr \rbrack \,.
\end{eqnarray}
The limit $\tilde Q\to 0$ in (\ref{Eqint3}) is still ill-defined since the
limits ${\bf K}_1-{\bf K}_2,\tilde Q\to 0$ do not commute in
(\ref{Eqint2}). Following Mermin, 
\cite{Mermin67,Lifshitz80} we first take the limit $\tilde Q\rightarrow 0$ 
(with either $Q/\Omega \rightarrow 0$ or $\Omega /Q\rightarrow 0$) and then 
${\bf K}_1-{\bf K}_2\rightarrow 0$: this ensures that $\Gamma ^Q$ and 
$\Gamma ^\Omega $ are continuous functions in the forward direction (${\bf K}
_1={\bf K}_2$). (Since $\Gamma ^Q$ and $\Gamma ^\Omega $ are ultimately 
connected to the (physical) forward scattering amplitude of two particles on 
the Fermi surface and to the Landau function $f_{\sigma _i}$, respectively, 
this requirement of continuity in the forward direction is very natural.) 
From (\ref{Eqint1},\ref{Eqint2}) we then conclude that the $Q$-limit 
of the forward scattering vertex function ($\Gamma ^Q$)  satisfies the 
Pauli principle while the $\Omega $-limit ($\Gamma ^\Omega $) does not, 
i.e.
\begin{eqnarray}
\Gamma ^Q_{\sigma _1\sigma _1,\sigma _2\sigma _3}
(\tilde K_1,\tilde K_1) &=&0 \,, \nonumber \\
\Gamma ^\Omega _{\sigma _1\sigma _1,\sigma _2\sigma _3}
(\tilde K_1,\tilde K_1) &\neq &0 \,. 
\label{Sym}
\end{eqnarray}

\subsubsection{Landau's solution}
\label{subsubsec:LS}

We first consider the functions $\Gamma ^{Q,\Omega }_{\sigma _i}(\tilde K_1,
\tilde K_2)$ and restrict ourselves to states on the Fermi surface 
($\omega =0$ and $K=K_F$). 
$\Gamma ^{Q,\Omega }_{\sigma _i}(\theta )$ become functions of the angle
$\theta $ between ${\bf K}_1$ and ${\bf K}_2$, and can be expanded on the
basis of Legendre polynomials (with coefficients $\Gamma ^{Q,\Omega
}_{\sigma _i}(l)$). The usual diagrammatic derivation of FLT does not take
into account the singularity which appears for $\tilde Q\to 0$ in the ZS$'$
channel when $\vert {\bf K}_1-{\bf K}_2\vert \ll T/v_F$. This can be
justified as follows.   
In general, physical quantities probe all the possible values of
the angle $\theta $. For example, the compressibility and the Pauli
susceptibility are entirely determined by $\Gamma _{\sigma _i}^Q(l=0)$. 
The singularity in the
ZS$'$ channel affects only small angles $\vert \theta \vert \ll T/E_F$
(where $E_F\sim v_FK_F$ is the Fermi energy), 
while the singularity in the ZS channel affects
all the angles. $\Gamma ^Q_{\sigma _i}(l)$ is therefore
determined by (\ref{Eqint1}) with an accuracy of order $T/E_F$ for any
reasonable value of $l$ ($l\ll E_F/T$). 

When (\ref{Eqint2}) is not taken into account, $\tilde \Gamma =\Gamma
^\Omega $ and (\ref{Eqint1}) becomes
\begin{equation}
\Gamma _{\sigma _i}^Q(l)=\Gamma _{\sigma _i}^\Omega (l)
-{{z^2N(0)} \over {2l+1}} \sum _{\sigma ,\sigma '}\Gamma _{\sigma _1\sigma
',\sigma \sigma _4} 
^\Omega (l)\Gamma ^Q_{\sigma \sigma _2,\sigma _3\sigma '}(l) \,.
\label{EQGL}
\end{equation}
If the spin dependent part of the 
interaction is due purely to exchange, one can write $\Gamma ^{Q,\Omega }
_{\sigma _i}$ as a function of a spin symmetric ($A^{Q,\Omega }$) and 
antisymmetric ($B^{Q,\Omega })$ part: 
\begin{equation}
2N(0)z^2\Gamma ^{Q,\Omega }_{\sigma _i}(l)=
A^{Q,\Omega }(l)
\delta _{\sigma _1,\sigma _4} \delta _{\sigma _2,\sigma _3}
+B^{Q,\Omega }(l)
\mbox {\boldmath $\tau $}_{\sigma _1\sigma _4} \cdot
\mbox {\boldmath $\tau $}_{\sigma _2\sigma _3} \,,
\label{AB}
\end{equation}
where $\mbox {\boldmath $\tau $}$ denotes the Pauli matrices. We then obtain 
from (\ref{EQGL})
 \begin{equation}
A^Q_l={{A^\Omega _l} \over {1+{{A^\Omega _l} \over {2l+1}}}} \,;\,\,\,\,\,
B^Q_l={{B^\Omega _l} \over {1+{{B^\Omega _l} \over {2l+1}}}} \,.
\label{ABres}
\end{equation}    

Eq.\ (\ref{Eqint1}) can also be used to obtain the collective modes (which
correspond to poles in the retarded vertex function) and any response
function at finite $\tilde Q$. One then recovers the results of the
phenomenological approach if one defines the Landau parameters by
$F^s_l=A^\Omega _l$ and $F^a_l=B^\Omega _l$, or equivalently
\begin{equation}
f_{\sigma _i}(\theta )= z^2 \Gamma ^\Omega _{\sigma _i}(\theta ) \,.
\label{Lfunction}
\end{equation}

Thus the microscopic FLT not only justifies the 
results obtained in the phenomenological approach but also provides a 
microscopic definition of the Landau parameters. It is important to 
note that
the Landau parameters do not correspond to a quantity entering the microscopic 
action or some low-energy effective action: it is necessary to integrate all 
the degrees of freedom to obtain $\Gamma ^\Omega _{\sigma _i}$ and therefore 
the Landau parameters. Generally, one does not try to calculate $\Gamma ^\Omega
_{\sigma _i}$ as a function of the microscopic parameters but only establishes 
the relation between this quantity and physical quantities which can be 
measured experimentally.

\section{RG equations at one-loop order}
\label{sec:1L}

From now on we restrict ourselves to a two-dimensional system since our 
discussion can be straightforwardly extended to the three-dimensional case. 
We consider interacting spin one-half fermions with a circular Fermi surface.
We write the partition function $Z$ as a functional integral 
over Grassmann variables,
\begin{equation}
Z=\int {\cal D}\psi ^* {\cal D}\psi e^{-S} \,,
\end{equation}
where, assuming that the high-energy degrees of freedom have been integrated 
out (in a functional sense), $S$ is a low-energy effective action describing 
the fermionic degrees of freedom with $\vert K-K_F \vert < \Lambda _0 \ll K_F$.
We write the effective action as
\begin{eqnarray}
S &=& -
\sum _{\tilde K,\sigma } \psi ^*_\sigma (\tilde K)(i\omega -\epsilon
({\bf K})) \psi _\sigma (\tilde K)
+{1 \over {4\beta \nu }} \sum _{\tilde K_1...\tilde K_4} \sum _{\sigma
_1... \sigma _4} U_{\sigma _1\sigma _2,\sigma _3\sigma _4} ({\bf K}_1,{\bf
K}_2,{\bf K}_3, {\bf K}_4)
\nonumber \\ & & \times  
\psi ^*_{\sigma _4}(\tilde K_4)
\psi ^*_{\sigma _3}(\tilde K_3)
\psi _{\sigma _2}(\tilde K_2)
\psi _{\sigma _1}(\tilde K_1)
\delta _{{\bf K}_1+{\bf K}_2,{\bf K}_3+{\bf K}_4}
\delta _{\omega _1+\omega _2,\omega _3+\omega _4} \,,
\label{action}
\end{eqnarray}
where the wave-vectors ${\bf K}$ satisfy $\vert K-K_F \vert < \Lambda _0$. 
$\beta =1/T$ is the inverse temperature and $\nu $ and $\tilde K$ have the same
meaning as in Sec.\ \ref{sec:FLT}. Ignoring irrelevant terms, we write the
single  
particle energy  as $\epsilon ({\bf K})=v_F(K-K_F)\equiv v_Fk$ (choosing the
chemical potential as the origin of the energies). The summation over the
wave-vectors is defined by
\begin{equation}
{1 \over \nu } \sum _{\bf K}=\int {{d^2{\bf K}} \over {(2\pi )^2}} 
\equiv K_F \int _{-\Lambda _0}^{\Lambda _0} {{dk} \over {2\pi }} 
\int _0^{2\pi } {{d\theta } \over {2\pi }} \,,
\end{equation}
ignoring irrelevant terms at tree-level. The coupling function $U_{\sigma
_1\sigma _2,\sigma _3\sigma _4} ({\bf K}_1,{\bf K}_2,{\bf K}_3, {\bf K}_4)$ 
is antisymmetric with respect to exchange of the two incoming or outgoing 
particles,
\begin{eqnarray}
U_{\sigma _1\sigma _2,\sigma _3\sigma _4} ({\bf K}_1,{\bf K}_2,{\bf K}_3, 
{\bf K}_4)&=&- U_{\sigma
_2\sigma _1,\sigma _3\sigma _4} ({\bf K}_2,{\bf K}_1,{\bf K}_3, {\bf K}_4)
\nonumber \\ &=& - U_{\sigma
_1\sigma _2,\sigma _4\sigma _3} ({\bf K}_1,{\bf K}_2,{\bf K}_4, {\bf K}_3) \,,
\label{Usym}
\end{eqnarray}
and is assumed to be a non-singular function of its arguments. 

The form of the action (\ref{action}) is usually justified by arguing that the
omitted terms are irrelevant according to tree-level analysis. 
\cite{Shankar94,Chitov95} This is not entirely correct. The
integration of high-energy modes ($\vert k\vert >\Lambda _0$) generates terms 
of order $n\geq 6$ in the $\psi ^{(*)}$ fields (i.e. three-, four-... -body
interactions) which are marginal although a naive
tree-level analysis would predict them to be irrelevant. The RG approach for
the low-energy modes ($\vert k\vert <\Lambda _0$) also produces such terms.
The role of these terms will be mentioned below and discussed in detail in
Sec.\ \ref{sec:afr}. Nevertheless, the form (\ref{action}) of the action is
sufficient for the purpose of this section.
Moreover, the possibility to assume that $U_{\sigma _i}$ is a regular
function of ${\bf K}_1,{\bf K}_2...$ and to ignore its dependence on 
the Matsubara frequencies $\omega _1$, $\omega _2$... (which is irrelevant at 
tree-level) is not obvious. Indeed, according to the results of the microscopic
FLT, we expect $U_{\sigma _i}$ to acquire singularities for small momentum and 
energy transfers. As will be shown below, these singularities arise in the
renormalization process when the 
momentum cut-off becomes smaller than $T/v_F$ so that they can be ignored in
the (bare) effective action (\ref{action}) if we choose $T\ll v_F\Lambda _0$. 
Note also that the integration of high-energy degrees of freedom generates a 
wave-function renormalization factor $z_{\Lambda _0}<1$ which has been 
eliminated from (\ref{action}) via a rescaling of the fermion fields. The $\psi
^{(*)}$'s in (\ref{action}) therefore do not correspond to the bare fermions 
but to quasi-particles with a renormalization factor $z_{\Lambda
_0}$.\cite{Nozieres64}  

As shown in Ref.\ \onlinecite{Shankar94}, the constraint
to have all momenta in the shell $\vert k\vert <\Lambda _0$ restricts the 
allowed scatterings to diffusion of particle-hole, or particle-particle, pairs
with small total momentum ($Q\lesssim \Lambda _0$). Consequently, only two 
coupling functions $U_{\sigma _i}$ have to be considered: the forward 
scattering coupling function and the BCS coupling function. 
In the following, we
neglect the latter by assuming it is irrelevant so that 
no BCS instability occurs. As in Sec.\ \ref{sec:FLT}, the forward 
scattering 
coupling function is denoted by $\Gamma _{\sigma _i}(\tilde K_1,\tilde K_2;
\tilde Q)$. According to tree-level analysis, this quantity is marginal and its
dependence on $k_{1,2}$ and $\omega _{1,2}$ is irrelevant. We therefore 
introduce  the coupling function 
$\Gamma _{\sigma _i}(\theta _1,\theta _2;\tilde Q)=
\Gamma _{\sigma _i}({\bf K}_1^F,{\bf K}_2^F;
\tilde Q)$ where ${\bf K}^F=K_F(\cos \theta ,\sin \theta )$ is a wave-vector
on the Fermi surface. It is not possible to put $\tilde Q=0$ in $\Gamma $
(although 
the dependence on $\tilde Q$ is irrelevant at tree-level) because $\Gamma $
will acquire a singular dependence on $\tilde Q$ in the process of
renormalization (this point is further discussed at the end of Sec.\
\ref{subsec:GQGO}). We decompose $\Gamma $  into a spin triplet
amplitude $\Gamma _t$ and a spin singlet amplitude $\Gamma _s$: 
\cite{Baym91,nota1}
\begin{equation}
 \Gamma _{\sigma _i}(\theta _1,\theta _2;\tilde Q) =
\Gamma _t  (\theta _1,\theta _2;\tilde Q) 
I^{\sigma _1\sigma _2}_{\sigma _3\sigma _4}
 + \Gamma _s  (\theta _1,\theta _2;\tilde Q) 
T^{\sigma _1\sigma _2}_{\sigma _3\sigma _4} \,,  
\end{equation}
where the functions 
\begin{eqnarray}
I^{\sigma _1\sigma _2}_{\sigma _3\sigma _4} &=& {1 \over 2} 
(\delta _{\sigma _1,\sigma _4}\delta _{\sigma _2,\sigma _3}+
\delta _{\sigma _1,\sigma _3}\delta _{\sigma _2,\sigma _4}) \,, \nonumber \\
T^{\sigma _1\sigma _2}_{\sigma _3\sigma _4} &=& {1 \over 2} 
(\delta _{\sigma _1,\sigma _4}\delta _{\sigma _2,\sigma _3}-
\delta _{\sigma _1,\sigma _3}\delta _{\sigma _2,\sigma _4}) 
\end{eqnarray}
satisfy the relations $I^{\sigma _1\sigma _2}_{\sigma _3\sigma _4}=
I^{\sigma _2\sigma _1}_{\sigma _3\sigma _4}
= I^{\sigma _1\sigma _2}_{\sigma _4\sigma _3}$,
$T^{\sigma _1\sigma _2}_{\sigma _3\sigma _4}=
-T^{\sigma _2\sigma _1}_{\sigma _3\sigma _4}
= -T^{\sigma _1\sigma _2}_{\sigma _4\sigma _3}$, and 
\begin{eqnarray}
I^{\sigma _1\sigma }_{\sigma _3\sigma '}
I^{\sigma '\sigma _2}_{\sigma \sigma _4}&=& 
{5 \over 4}I^{\sigma _1\sigma _2}_{\sigma _3\sigma _4}  
-{3 \over 4} T^{\sigma _1\sigma _2}_{\sigma _3\sigma _4} \,, \nonumber \\
I^{\sigma _1\sigma }_{\sigma _3\sigma '}
T^{\sigma '\sigma _2}_{\sigma \sigma _4}&=& 
-{1 \over 4}I^{\sigma _1\sigma _2}_{\sigma _3\sigma _4}  
+{3 \over 4} T^{\sigma _1\sigma _2}_{\sigma _3\sigma _4} \,, \nonumber \\
T^{\sigma _1\sigma }_{\sigma _3\sigma '}
T^{\sigma '\sigma _2}_{\sigma \sigma _4}&=& 
{1 \over 4}I^{\sigma _1\sigma _2}_{\sigma _3\sigma _4}  
+{1 \over 4} T^{\sigma _1\sigma _2}_{\sigma _3\sigma _4} \,,
\label{propIT}
\end{eqnarray}
where a sum over $\sigma $ and $\sigma '$ is implied.

The KW RG procedure consists in successive partial 
integrations of
the fermion field degrees of freedom in the infinitesimal momentum shell
$\Lambda _0e^{-dt}\leq \vert k\vert \leq \Lambda _0$ where $dt$ is the 
RG generator and $\Lambda (t)=\Lambda
_0e^{-t}$ the effective momentum cut-off at step $t$ . Each
partial integration is followed by a rescaling of radial momenta,
frequencies and fields
(i.e $\omega '=s\omega $, $k'=sk$ and ${\psi ^{(*)}}'=\psi ^{(*)}$ with
$s=e^{dt}$) in order to let the quadratic part of the action
(\ref{action}) invariant
and to restore the initial value of the cut-off. (See Refs.\  
\onlinecite{Bourbon91,Shankar94} for a detailed presentation of the KW RG
method 
applied to fermion systems.) The partial integration, which is evaluated 
perturbatively within the framework of a loop expansion, modifies the 
parameters of the action which become functions of the flow parameter $t$.
It also generates higher order interactions (three-, four-... body
interactions) whose relevance or irrelevance should be
controlled (see Sec.\ \ref{sec:afr}). \cite{Bourbon91} 

At one-loop order, the three diagrams which have to be 
considered for the renormalization of $\Gamma _{\sigma _i}(\theta _1,\theta
_2;\tilde Q)$ are shown in Fig.\ 
\ref{fig:corr_1loop}. In these diagrams, the momenta of the internal
lines should be in the infinitesimal shell which is integrated out.

\subsection{RG equations for $\Gamma ^Q_{\sigma _i}$ and $\Gamma ^\Omega
_{\sigma _i}$} 
\label{subsec:GQGO}

We first consider the RG equations for the $Q$- and $\Omega $-limits of the
forward scattering coupling function: 
\begin{eqnarray}
\Gamma _{\sigma _i}^Q(\theta _1-\theta _2) &=& \lim _{Q\rightarrow 0} \Bigl 
\lbrack \Gamma _{\sigma _i}(\theta _1,\theta _2;\tilde Q) 
\Bigl \vert _{\Omega =0} \Bigr \rbrack \,, \nonumber \\
\Gamma _{\sigma _i}^\Omega (\theta _1-\theta _2) &=& \lim _{\Omega 
\rightarrow 0} \Bigl  \lbrack \Gamma _{\sigma _i}(\theta _1,\theta _2;\tilde
Q) \Bigl \vert _{Q=0} \Bigr \rbrack \,, 
\end{eqnarray}
since these two quantities play an important role in the microscopic FLT. 
$\Gamma _{\sigma _i}^{Q,\Omega }(\theta )$ are even functions of $\theta $.  
The only remnant of the
antisymmetry of $U_{\sigma _i}$ (Eq.\ (\ref{Usym})) is the condition
\cite{Chitov95}
\begin{equation}
\Gamma ^Q_t(\theta =0)\vert _{\Lambda (t)=\Lambda _0} =\Gamma
^\Omega _t(\theta =0)\vert _{\Lambda (t)=\Lambda _0}=0 \,,
\end{equation}
using the fact that $U_{\sigma _i}$ is assumed to be a regular function of
its arguments. 

We ignore the symmetry-preserving contribution of the BCS channel (see
however Sec.\ \ref{subsec:ibc})  and first discuss the contribution of 
the ZS graph. This graph involves the quantity
\begin{eqnarray}
T\sum _{\omega } G(\tilde K)G(\tilde K+\tilde Q) &=& {1 \over 2}
{{\tanh \Bigl \lbrack {\beta \over 2}\epsilon ({\bf K}+{\bf Q}) \Bigr \rbrack 
- \tanh \Bigl \lbrack {\beta \over
2}\epsilon ({\bf K}) \Bigr \rbrack }  \over
{i\Omega +\epsilon ({\bf K})-\epsilon ({\bf K}
+{\bf Q})}} \nonumber \\
&\simeq & {{v_F{\bf \hat K}\cdot {\bf Q}} 
\over {i\Omega -v_F{\bf \hat K}\cdot {\bf Q}}}
{{\beta } \over {4\cosh ^2(\beta v_Fk/2)}} 
\label{bubbleZS}
\end{eqnarray}
for small $Q$. Here $G(\tilde K)=(i\omega -v_Fk)^{-1}$ is the 
one-particle Green's function and $\vert k\vert =\Lambda (t)$. 
It is clear that the limit of (\ref{bubbleZS}) for 
$\tilde Q\rightarrow 0$ depends on the ratio $Q/\Omega $. It equals $-(\beta
/4) \cosh ^{-2}(\beta v_Fk/2)$ in the $Q$-limit while it vanishes 
in the $\Omega $-limit. We obtain
\begin{eqnarray}
{{d\Gamma ^Q_{\sigma _i}(\theta _1-\theta _2)} \over {dt}} \Biggl \vert
_{\rm ZS}  &=& - {{N(0)\beta _R} \over 
{\cosh ^2(\beta _R)}} \int {{d\theta } \over {2\pi }} 
\sum _{\sigma ,\sigma '} 
\Gamma _{\sigma _1\sigma ',\sigma \sigma _4}^Q (\theta _1-\theta )
\Gamma _{\sigma \sigma _2,\sigma _3\sigma '}^Q (\theta -\theta _2) \,,
\nonumber \\
{{d\Gamma ^\Omega _{\sigma _i}(\theta _1-\theta _2)}  \over {dt}} 
\Biggl \vert _{\rm ZS} &=& 0 \,,
\label{ZS1}
\end{eqnarray}
a result which was first obtained in Ref.\ \onlinecite{Chitov95}.  
Here $\beta _R=v_F\beta \Lambda (t)/2$ is a dimensionless inverse 
temperature and $N(0)=K_F/2\pi v_F$ the density of states per spin. 
Note that we have obtained (\ref{ZS1}) without rescaling the
radial momenta, frequencies and fields. However, the same result is obtained
if one chooses to do this rescaling. In this case, $T$ should be replaced by
$T(t)=Te^t$ (this $t$-dependence follows from the rescaling of the Matsubara
frequencies) and $\vert k\vert =\Lambda _0$ in (\ref{bubbleZS}), which leads
to (\ref{ZS1}) with $\beta _R=v_F\beta (t)\Lambda _0/2$ (which can be also
written $\beta _R=v_F\beta \Lambda (t)/2$ with $\Lambda (t)=\Lambda
_0e^{-t}$). Since the two procedures (with
or without rescaling) are equivalent, we will in general use $\beta
=1/T$ and $\Lambda (t)=\Lambda _0e^{-t}$ which amounts to keeping the original
units for the frequencies and momenta. Eqs.\ (\ref{ZS1}) show that
we recover two important results of the microscopic
FLT: i) the contribution of the ZS graph for $\tilde Q\to 0$ strongly
depends on the order of
the limits $Q\to 0$ and $\Omega \to 0$. ii) this singularity at $\tilde Q\to 0$
comes from the integration of states near the Fermi surface ($\Lambda (t)
\lesssim T/v_F$) since $\beta _R/\cosh ^2(\beta _R)\ll 1$ for $\beta _R\gg 1$. 
Since $\lim _{\beta \rightarrow \infty } (\beta /4)\cosh ^{-2}(\beta
x/2)=\delta (x)$, the ZS graph gives a singular contribution (with respect
to $\Lambda (t)$) to the RG flow of $\Gamma ^Q$ when $T\rightarrow 0$.

The ZS$'$ graph involves the quantity
\begin{equation}
T\sum _\omega G(\tilde K)G(\tilde K+{\bf K}_{21}^F-\tilde Q) 
= {1 \over 2} 
{{\tanh \Bigl \lbrack {\beta \over 2}\epsilon ({\bf K}+{\bf K}_{21}^F
-{\bf Q}) \Bigr \rbrack  - \tanh \Bigl \lbrack {\beta \over
2}\epsilon ({\bf K}) \Bigr \rbrack }  \over
{-i\Omega +\epsilon ({\bf K})-\epsilon ({\bf K}
+{\bf K}_{21}^F-{\bf Q})}}   \,,
\label{ZS'1}
\end{equation}
where ${\bf K}_{21}^F={\bf K}_2^F-{\bf K}_1^F$.
Since both internal lines should have their momenta in the infinitesimal shell 
near the cut-off $\Lambda (t)$, the ZS$'$ graph is non zero in the limit 
$Q\rightarrow 0$ only if ${\bf K}_{21}^F\rightarrow 0$. We
therefore obtain a (non-physical) discontinuity in the forward direction
($\theta =0$). As discussed in Sec.\ \ref{sec:afr}, it is necessary to
consider three-body interactions between fermions to restore the continuity
in the forward direction. The ZS$'$ graph with ${\bf K}_{21}^F \neq
0$ is then generated from the three-particle vertex function. 
To calculate the ZS$'$ graph for small but finite $\vert {\bf K}_{21} 
\vert $, we adopt in this section the following approximate procedure
which is 
justified in Sec.\ \ref{sec:afr}. We impose ${\bf K}$ to be at the cut-off
but relax the constraint to have ${\bf K}+{\bf K}_{21}^F -{\bf Q}$
at the cut-off. For small ${\bf K}_{21}^F$ (i.e. for  $\vert
\theta _1-\theta _2 \vert \ll T/E_F$) and small $Q$, (\ref{ZS'1}) becomes
\begin{equation}
{{v_F{\bf \hat K}\cdot ({\bf K}_{21}^F-{\bf Q})} 
\over {-i\Omega -v_F{\bf \hat K}\cdot ({\bf K}_{21}^F-{\bf Q})}}
{{\beta } \over {4\cosh ^2(\beta v_Fk/2)}} \,.
\label{ZS'2}
\end{equation}
It is clear from this expression that the limits $\tilde Q,{\bf K}_{21}^F
\rightarrow 0$ do not commute. The same problem arises in the 
microscopic FLT as pointed out by Mermin \cite{Mermin67,Lifshitz80,Hewson94}
and briefly
discussed in Sec.\ \ref{sec:FLT}. In agreement with the procedure followed
in the microscopic FLT, we first take the
limit $\tilde Q\rightarrow 0$ (which is well defined for ${\bf K}_{21}^F 
\neq 0$) and then $\theta _1-\theta _2\rightarrow 0$. This procedure
(together with the consideration of three-body interactions) ensures the
continuity of $\Gamma ^{Q,\Omega }(\theta )$ in the forward direction. We have
\begin{equation} 
\lim _{\theta _1-\theta _2\rightarrow 0} \Bigl \lbrack
T\sum _\omega G(\tilde K)G(\tilde K+{\bf K}_{21}^F-\tilde Q)
\Bigl \vert _{\tilde Q=0} \Bigr \rbrack =
-{\beta \over {4\cosh ^2(\beta v_Fk/2)}} \,.
\end{equation}
At low temperature, the ZS$'$ graph also gives a singular contribution (with
respect to $\Lambda (t)$) to the flows of $\Gamma ^Q$
and $\Gamma ^\Omega $ when $\theta _1-\theta _2\rightarrow 0$. 
The singularity in the ZS$'$
channel is restricted to the angles $\vert \theta _1-\theta _2 \vert \ll T
/E_F$. For $\vert \theta _1-\theta _2 \vert \gg T/E_F$, the ZS$'$ channel gives
a smooth contribution to the flow of $\Gamma ^{Q,\Omega }$. Note also that as 
in the microscopic FLT the ZS$'$ graph does not differentiate between $\Gamma 
^Q$ and $\Gamma ^\Omega $ contrary to the ZS graph. 

Taking into account the spin dependence of the coupling functions (using 
(\ref{propIT})), we obtain that the
contributions of the ZS and ZS$'$ graphs to the RG flow of $\Gamma ^Q_t(\theta
=0)$ cancel each other:
\begin{equation}
{{d\Gamma ^Q_t(\theta =0)} \over {dt}} \Biggl \vert _{{\rm ZS}'} =
-{{d\Gamma ^Q_t(\theta =0)} \over {dt}} \Biggl \vert _{\rm ZS} \,.
\end{equation}
Consequently, we have $\Gamma ^Q_t(\theta =0)=0$ for any value of the flow 
parameter $t$. The antisymmetry of $\Gamma ^Q$ is therefore conserved under RG.
On the other hand, since the contribution of the ZS graph to the flow of
$\Gamma ^\Omega (\theta =0)$ vanishes, while the contribution of the ZS$'$ 
graph does not, the antisymmetry of $\Gamma ^\Omega $ is not conserved under 
RG. The symmetry properties of the two-particle vertex function agree
with the results of the microscopic FLT (Eqs.\ (\ref{Sym})). The antisymmetry
of $\Gamma ^\Omega $ is lost only when $\Lambda (t)\lesssim T/v_F$. For 
$\Lambda (t)\gg T/v_F$, the RG flow is determined by the ZS$'$ channel only and
$\Gamma ^Q(\theta )\simeq \Gamma ^\Omega (\theta )$. 

Taking into account both the contributions of the ZS and ZS$'$ graphs, the RG
equations of $\Gamma ^{Q,\Omega }$ can be written as
\begin{eqnarray}
{{d\Gamma ^Q_{\sigma _i}} \over {dt}} &=& 
{{d\Gamma ^Q_{\sigma _i}} \over {dt}} \Biggl \vert _{\rm ZS} +
{{d\Gamma ^Q_{\sigma _i}} \over {dt}} \Biggl \vert _{{\rm ZS}'} \,,
\nonumber \\ 
{{d\Gamma ^\Omega _{\sigma _i}} \over {dt}} &=& 
{{d\Gamma ^\Omega _{\sigma _i}} \over {dt}} \Biggl \vert _{{\rm ZS}'} =
{{d\Gamma ^Q_{\sigma _i}} \over {dt}} \Biggl \vert _{{\rm ZS}'} \,. 
\end{eqnarray}
The two preceding equations can be combined (using also (\ref{ZS1})) to 
obtain an equation relating $\Gamma ^Q$ and $\Gamma ^\Omega $:
\begin{eqnarray}
{{d\Gamma ^Q_{\sigma _i}(\theta _1-\theta _2)} \over {dt}} &=& 
{{d\Gamma ^\Omega _{\sigma _i}(\theta _1-\theta _2)} \over {dt}}
 \nonumber \\
&& - {{N(0)\beta _R} \over 
{\cosh ^2(\beta _R)}}  
\int {{d\theta } \over {2\pi }} 
\sum _{\sigma ,\sigma '} 
\Gamma _{\sigma _1\sigma ',\sigma \sigma _4}^Q (\theta _1-\theta )
\Gamma _{\sigma \sigma _2,\sigma _3\sigma '}^Q (\theta -\theta _2) \,.
\label{RGEQ1}
\end{eqnarray}
Introducing the Fourier transforms
\begin{equation}
\Gamma ^{Q,\Omega }_{\sigma _i}(l) = \int {{d\theta } \over {2\pi }}
e^{-il\theta } \Gamma ^{Q,\Omega }_{\sigma _i}(\theta ) \,,
\end{equation}
and performing the sum over spins using (\ref{propIT}), we obtain 
\begin{eqnarray}
{{d\Gamma ^Q_t(l)} \over {dt}} &=& -{{N(0)\beta _R} \over {\cosh ^2(\beta _R)}}
\Biggl \lbrack {5 \over 4} {\Gamma ^Q_t(l)}^2 +{1 \over 2} 
\Gamma ^Q_t(l) \Gamma ^Q_s(l) +{1 \over 4} {\Gamma ^Q_s(l)}^2 \Biggr \rbrack 
+ {{d\Gamma ^\Omega _t(l)} \over {dt}} \,,  \nonumber \\
{{d\Gamma ^Q_s(l)} \over {dt}} &=& -{{N(0)\beta _R} \over {\cosh ^2(\beta _R)}}
\Biggl \lbrack {3 \over 4} {\Gamma ^Q_t(l)}^2 +{3 \over 2} 
\Gamma ^Q_t(l) \Gamma ^Q_s(l) -{1 \over 4} {\Gamma ^Q_s(l)}^2 \Biggr \rbrack 
+ {{d\Gamma ^\Omega _s(l)} \over {dt}} \,. 
\end{eqnarray}
These equations agree with those of Chitov and S\'en\'echal apart from the
terms $d\Gamma ^\Omega _{t,s}/dt$ coming from the ZS$'$ graph which were 
omitted in Ref.\ \onlinecite{Chitov95}.
Introducing the spin symmetric ($A^{Q,\Omega }$) and antisymmetric ($B^{Q,
\Omega }$) parts as in Eq.\ (\ref{AB}), and using
\begin{eqnarray}
2N(0)\Gamma ^{Q,\Omega }_t(l)&=&A^{Q,\Omega }(l)+B^{Q,\Omega }(l) \,,
\nonumber \\
2N(0)\Gamma ^{Q,\Omega }_s(l)&=&A^{Q,\Omega }(l)-3B^{Q,\Omega }(l) \,,
\label{AB1}
\end{eqnarray}
the RG equations take the simple form
\begin{eqnarray}
{{dA^Q_l} \over {dt}} &=& {{dA^\Omega _l} \over {dt}} 
-{{\beta _R} \over {\cosh ^2(\beta _R)}} {A^Q_l}^2 \,, \nonumber \\
{{dB^Q_l} \over {dt}} &=& {{dB^\Omega _l} \over {dt}} 
-{{\beta _R} \over {\cosh ^2(\beta _R)}} {B^Q_l}^2 \,.
\label{ABeq}
\end{eqnarray}
The two preceding equations are exact at one-loop order. In order to solve them
approximately in the low temperature limit, we will take advantage of the 
singularities which arise in the RG flow. We integrate Eqs.\ (\ref{ABeq}) 
between 0 and $t$ to obtain (writing explicitly the $t$ dependence)
\begin{equation}
A^Q_l(t)=A^\Omega _l(t) - \int _0^t dt' \, 
{{\beta _R} \over {\cosh ^2(\beta _R)}} A^Q_l(t')^2 
\label{RGEQ2}
\end{equation}
and a similar equation for $B^Q_l(t)$. Here we have used $A^Q_l(t=0)=A^\Omega 
_l(t=0)$ since $\Gamma _{\sigma _i}(t=0)=U_{\sigma _i}$ is a non-singular 
function of its arguments as was justified above when $T\ll v_F\Lambda _0$. 
Iterating (\ref{RGEQ2}), we obtain
\begin{equation}
A^Q_l(t)=A^\Omega _l(t) - \int _0^t dt' \, 
{{\beta _R} \over {\cosh ^2(\beta _R)}} A^\Omega _l(t')^2 +...
\label{RGEQ21}
\end{equation}
We have shown above that $\Gamma _{\sigma _i}^\Omega (\theta )$ is a smooth
function of $\Lambda (t)$ except for small angles $\vert \theta \vert \ll
T/v_F$. The Fourier transform  $\Gamma _{\sigma _i}^\Omega (l)$ is then a
smooth function of $\Lambda (t)$ (for $l\ll E_F/T$). At low temperature, we
can therefore make the approximation $A^\Omega _l(t)\vert _{\Lambda
(t)\sim T/v_F}\simeq {A^\Omega _l}^*$ where ${A^\Omega _l}^*=A^\Omega _l
\vert _{\Lambda (t)=0}$ is the fixed point (FP) value of $A^\Omega _l$. 
Since the thermal
factor $\beta _R/\cosh ^2(\beta _R)$ 
is strongly peaked for $\Lambda (t')\lesssim T/v_F$, we can replace
$A^\Omega _l(t')$ in the rhs of (\ref{RGEQ21}) by 
${A^\Omega _l}^*$. For $\Lambda
(t)\lesssim T/v_F$, (\ref{RGEQ21}) then becomes 
\begin{equation}
A^Q_l(t)={A^\Omega _l}^*-\int _0^t dt' \, 
{{\beta _R} \over {\cosh ^2(\beta _R)}} A^Q_l(t')^2 \,.
\label{RGEQ3}
\end{equation}
The preceding equation shows that the ZS channel is now decoupled from the
other channels. As in the microscopic FLT, this property follows from the
fact that the singularity of the ZS channel is due to particle-hole pairs
close to the Fermi surface ($\Lambda (t)\lesssim T/v_F$).
Eq.\  (\ref{RGEQ3}) is solved by introducing the parameter $\tau =\tanh
\beta _R$, which leads to 
\begin{eqnarray}
A^Q_l(\tau ) &=& {{{A^\Omega _l}^*} \over {1+(\tau _0-\tau ){A^\Omega _l}^*}} 
\,, \nonumber \\
B^Q_l(\tau ) &=& {{{B^\Omega _l}^*} \over {1+(\tau _0-\tau ){B^\Omega _l}^*}} 
\,, \label{ABflow}
\end{eqnarray}
for $\Lambda (t)\lesssim T/v_F$. 
$\tau _0=\tanh (\beta v_F\Lambda _0/2)\simeq 1$ for $T\ll v_F\Lambda _0$. 
Eqs.\ (\ref{ABflow}) show that the
marginality of the coupling functions is lost at one loop-order. $A^Q_l$ and
$B^Q_l$ are either relevant or irrelevant depending on their sign. For
${A^\Omega _l}^*<-1$ (${B^\Omega _l}^*<-1$), $A^Q_l$ ($B^Q_l$) diverges
for some value of $\tau $ which signals an instability of the Fermi liquid.
For instance, if ${B^\Omega _0}^*<-1$, the Fermi liquid is unstable with
respect to a ferromagnetic phase. 
(In (\ref{ABres}), one should  replace $2l+1$ by 1 for a two-dimensional
system.) The stability conditions ${A^\Omega _l}^*>-1$ and ${B^\Omega _l}^*
>-1$ are known as the Pomeranchuk's stability conditions. \cite{Lifshitz80} 
The FP values of $A^Q_l$ and $B^Q_l$
are obtained for $\tau =0$ ($\Lambda (\tau =0)=0$):
\begin{equation}
{A^Q_l}^*= {{{A^\Omega _l}^*} \over {1+{A^\Omega _l}^*}}\,;
\,\,\,\,
 {B^Q_l}^*= {{{B^\Omega _l}^*} \over {1+{B^\Omega _l}^*}}\,. 
\label{PV1}
\end{equation} 
Here and in the following we put $\tau _0=1$.  

The RG at one-loop order therefore agrees with the microscopic FLT
(Eq.\ (\ref{ABres}) with $2l+1$ replaced by 1).
Eq.\ (\ref{RGEQ3}) (together with the analog equation for $B^Q_l$)
is nothing else but a Bethe-Salpeter equation in the ZS channel for the
vertex $\Gamma _{\sigma _i}^Q$, with ${\Gamma ^\Omega }^*$ the irreducible
two-particle vertex function. This shows that the integration of the RG
equations generates the same diagrams as those considered by 
Landau \cite{Landau59,Abrikosov63,Nozieres64} (as
shown in Sec.\ \ref{sec:bol}, Eq.\ (\ref{RGEQ3}) holds at all orders in a
loop expansion). From this point of
view, there is therefore a strict equivalence between the RG approach and
the microscopic FLT. 

It has 
been claimed in Ref.\ \onlinecite{Shankar94} that the coupling function
$\Gamma  
_{\sigma _i}$ cannot become singular because the integration of an 
infinitesimal momentum shell cannot produce any non-analyticity. This 
argument is not correct since non-analyticity can originate in the infinite sum
over the Matsubara frequencies. In this section, we have obtained a 
non-analyticity summing the product $G(\tilde K)G(\tilde K+\tilde Q)$ over 
$\omega $. Because of this non-analyticity, one should 
keep in the action all coupling functions $\Gamma (\theta _1,\theta _2;\tilde Q
)$ whatever the value of $\tilde Q$ (with $Q\lesssim \Lambda (t)$). To 
illustrate this point, consider a marginal 
variable (call it $g$) which is a function of $\tilde Q$. Since $g$ is 
marginal, it is tempting to neglect its dependence on $\tilde Q$ arguing it is 
irrelevant. This latter point is usually proved by considering the Taylor 
expansion of $g(\tilde Q)$:
\begin{equation}
g(\tilde Q)=g_{00}+g_{10}Q+g_{01}\Omega +g_{11}Q\Omega + \cdot \cdot \cdot \,.
\end{equation}
A dimensional analysis shows that $g_{00}$ is marginal and all other 
coefficients irrelevant, so that only $g_{00}$ has to be kept in the action. 
This argument is correct only if $g$ is a regular function at $\tilde Q=0$. 
Otherwise, it has no Taylor expansion around $\tilde Q=0$ and there is no way 
to control the marginality or irrelevance of the dependence on $\tilde Q$. 

The importance of keeping the full dependence of $\Gamma (\theta _1,\theta
_2;\tilde Q)$ on $\tilde Q$ can be understood from a more physical
argument. A neutral Fermi liquid with short 
range interaction can sustain a collective charge (or spin) density oscillation
with an excitation energy vanishing in the limit of long-wave length (as
shown by Mermin, \cite{Mermin67} there exists at least one such mode). This 
collective mode strongly affects $\Gamma ^*(\theta _1,\theta _2;\tilde Q)$
since it yields a pole at $\Omega =c_0Q$ ($c_0$ is the velocity of the 
zero-sound or spin-waves mode) in the retarded two-particle vertex function
obtained by analytical continuation $i\Omega \rightarrow \Omega
+i0^+$ (see Sec.\ \ref{subsec:cm}). This shows that the dependence on $\tilde
Q$ cannot be irrelevant.

\subsection{RG equations for $\Gamma _{\sigma _i}(\theta _1,\theta _2;\tilde
Q)$: collective modes}
\label{subsec:cm}

RG equations for $\Gamma _{\sigma _i}(\theta _1,\theta _2;\tilde Q)$ with
$\tilde Q\neq 0$ are a priori difficult to obtain, since in general the 
one-loop graphs vanish when $Q\neq 0$. For example, it is not possible to have
both ${\bf K}$ and ${\bf K}+{\bf Q}$ in the infinitesimal momentum
shell to be integrated out when ${\bf Q}\neq 0$ (except for very rare
configurations) so that the ZS graph vanishes. The
same kind of problem arises in the calculation of the ZS$'$ graph even when
$\tilde Q\rightarrow 0$ (see preceding section). As discussed in Sec.\
\ref{sec:afr}, a one-loop calculation should involve the consideration of
three-body interactions. In this section, we restrict ourselves to the
case of finite but small $\tilde Q$. We adopt the approximate procedure used
in Sec. \ref{subsec:GQGO} for the calculation of the ZS$'$ graph: we impose on
${\bf K}$ to be in the infinitesimal momentum shell to be integrated out 
but relax the analog condition for $\bf K+\bf Q$. As shown below,
this approximate procedure 
is sufficient (and correct) to obtain the long wave-length limit of $\Gamma
_{\sigma _i}(\theta _1,\theta _2;\tilde Q)$.

According to Sec.\ \ref{subsec:GQGO}, the ZS$'$ graph does not produce any
singularity when $\tilde Q\rightarrow 0$. We can safely put
$\tilde Q=0$ in the contribution of this graph to $\Gamma $. The dependence
on $\tilde Q$ of the ZS graph is given by (\ref{bubbleZS}). The RG equation
(\ref{RGEQ1}) then becomes:
\begin{eqnarray}
{{d\Gamma _{\sigma _i}(\theta _1,\theta _2;\tilde Q)} \over {dt}} = 
{{d\Gamma ^\Omega _{\sigma _i}(\theta _1-\theta _2)} \over {dt}}
&+& {{N(0)\beta _R} \over 
{\cosh ^2(\beta _R)}}  
\int {{d\theta } \over {2\pi }} g(\theta ,\tilde Q)
\nonumber \\ && \times
\sum _{\sigma ,\sigma '} 
\Gamma _{\sigma _1\sigma ',\sigma \sigma _4} (\theta _1,\theta ;\tilde Q)
\Gamma _{\sigma \sigma _2,\sigma _3\sigma '} (\theta ,\theta _2;\tilde Q) \,,
\label{RGEQ4}
\end{eqnarray}
where $g(\theta ,\tilde Q)=v_F{\bf \hat K}\cdot {\bf Q} /
(i\Omega -v_F{\bf \hat K}\cdot {\bf Q})$ and 
${\bf \hat K}=(\cos \theta ,\sin \theta )$. $\Gamma ^\Omega _{\sigma
_i}(\theta _1-\theta _2)$ is a smooth function of $\Lambda (t)$ except for
$\vert \theta _1-\theta _2\vert \ll T/E_F$. If we are interested in
quantities (like the collective modes) which involve all the values of
$\theta _1-\theta _2$, we can consider $\Gamma ^\Omega$ as a smooth
function. It is then possible to proceed as in Sec.\ \ref{subsec:GQGO}
to solve (\ref{RGEQ4}). We introduce the quantities $A(\theta _1,\theta
_2;\tilde Q)$ and $B(\theta _1,\theta _2;\tilde Q)$ by analogy with
(\ref{AB1}). Eq.\ (\ref{RGEQ4}) becomes
\begin{eqnarray}
{{dA(\theta _1,\theta _2;\tilde Q)} \over {dt}} &=& 
{{dA^\Omega (\theta _1-\theta _2)} \over {dt}} 
+{{\beta _R} \over {\cosh ^2(\beta _R)}} \int {{d\theta } \over {2\pi }}
g(\theta ,\tilde Q)
A(\theta _1,\theta ;\tilde Q)A(\theta ,\theta _2;\tilde Q) \,, \nonumber \\
{{dB(\theta _1,\theta _2;\tilde Q)} \over {dt}} &=& 
{{dB^\Omega (\theta _1-\theta _2)} \over {dt}} 
+{{\beta _R} \over {\cosh ^2(\beta _R)}} \int {{d\theta } \over {2\pi }}
g(\theta ,\tilde Q)
B(\theta _1,\theta ;\tilde Q)B(\theta ,\theta _2;\tilde Q) \,.
\end{eqnarray}
Integrating this equation, and taking advantage of the singularity of $\beta
_R/\cosh ^2(\beta _R)$ when $T\rightarrow 0$, we obtain for $\Lambda
(t)\lesssim T/v_F$ 
\begin{equation}
A(\theta _1,\theta _2;\tilde Q)=
{A^\Omega }^*(\theta _1-\theta _2)
-\int {{d\theta } \over {2\pi }} 
g(\theta ,\tilde Q)
\int _{1}^\tau d\tau ' 
A(\theta _1,\theta ;\tilde Q) A(\theta ,\theta _2;\tilde Q)
\,,
\end{equation}
and a similar equation for $B$. If we iterate this equation, we obtain:
\begin{eqnarray}
A(\theta _1,\theta _2;\tilde Q)&=&
{A^\Omega }^*(\theta _1-\theta _2) +(1-\tau )\int {{d\theta } \over
{2\pi }} g(\theta ,\tilde Q) {A^\Omega }^*(\theta _1-\theta ) {A^\Omega
}^*(\theta -\theta _2)   
\nonumber \\ && + (1-\tau )^2 \int {{d\theta } \over {2\pi }}
g(\theta ,\tilde Q)
\int {{d\theta '} \over {2\pi }} g(\theta ',\tilde Q)
{A^\Omega }^*(\theta _1-\theta ) {A^\Omega }^*(\theta -\theta ') 
{A^\Omega }^*(\theta '-\theta _2) +\cdot \cdot \cdot
\end{eqnarray}
This expansion is clearly equivalent to the integral equation
\begin{eqnarray}
A(\theta _1,\theta _2;\tilde Q)&=&
{A^\Omega }^*(\theta _1-\theta _2) + (1-\tau )
\int {{d\theta } \over {2\pi }}
g(\theta ,\tilde Q)
{A^\Omega }^*(\theta _1-\theta ) A(\theta ,\theta _2;\tilde Q) \,, \nonumber
\\
B(\theta _1,\theta _2;\tilde Q)&=&
{B^\Omega }^*(\theta _1-\theta _2) + (1-\tau )
\int {{d\theta } \over {2\pi }} g(\theta ,\tilde Q)
{B^\Omega }^*(\theta _1-\theta ) B(\theta ,\theta _2;\tilde Q) \,.
\label{RGEQ5}
\end{eqnarray}
At the fixed point ($\tau =0$), we recover the equations which determine
the two-particle vertex function in the microscopic FLT (Eq.\ (\ref{Eqint1})
evaluated for states on the Fermi surface). In the above calculation, the
dependence on $\tilde Q$ follows from (\ref{bubbleZS}) which is correct only
in the limit $Q\to 0$. It is clear that for finite $Q$, the singularity
$(\beta /4)\cosh ^{-2}(\beta v_Fk/2)\vert _{T\to 0}=\delta (v_Fk)$ is
weakened (only the flow of $\Gamma ^Q$ presents a singularity $\sim \delta
(\Lambda )$ for $T\to 0$). Consequently, the determination of $\Gamma
^*_{\sigma _i}(\theta _1,\theta _2;\tilde Q)$ is less accurate at finite
$Q$.  

The spectrum of the collective
modes is given by the poles of $A$ (zero-sound mode) and $B$ (spin-waves
mode) after analytical continuation $i\Omega \rightarrow \Omega +i0^+$. As
in the microscopic FLT, we define the Landau parameters in such a way that
the results of the phenomenological approach are reproduced. We therefore
identify
the Landau parameters with the FP values of $A^\Omega _l$ and $B^\Omega _l$,
$F^s_l={A^\Omega _l}^*$ and $F^a_l={B^\Omega _l}^*$, or equivalently
$f_{\sigma _i}(\theta )={\Gamma ^\Omega _{\sigma _i}}^*(\theta )$.
Considering
the fact that the $\psi ^{(*)}$'s in (\ref{action}) have been rescaled to 
eliminate the wave-function renormalization factor $z_{\Lambda _0}$, we 
eventually come to the following definition of the Landau function:
\begin{equation}
f_{\sigma _i}(\theta )=z_{\Lambda _0}^2
{\Gamma ^\Omega _{\sigma _i}}^*(\theta ) \,,
\end{equation}
where ${\Gamma ^\Omega _{\sigma _i}}^*(\theta )$ now refers to the bare
fermions. 

In the following, we will sometimes consider the simple case where
$F^{s,a}_l=0$ if $l\neq 0$. Eqs.\ (\ref{RGEQ5}) then yield
\begin{eqnarray}
A(\tilde Q) &=& {{F^s_0} \over {1+(1-\tau )F^s_0\Omega _0(\eta )}}
\,, \nonumber \\
B(\tilde Q) &=& {{F^a_0} \over {1+(1-\tau )F^a_0\Omega _0(\eta )}}
\,,  
\label{AB00}
\end{eqnarray}
where $\eta =i\Omega /v_FQ$ and
\begin{equation}
\Omega _0(x)=\int {{d\theta } \over {2\pi }} {{\cos \theta }
\over {\cos \theta - x}} \,.
\end{equation}

\subsection{Density-density response function}
\label{subsec:ddrf}

We show in this section how the density-density response function can be
calculated in the KW RG approach. We proceed as in Ref.\
\onlinecite{Bourbon91}. We introduce a source term in the action, 
\begin{equation}
S_h=- \sum _{\tilde Q} h(\tilde Q) \rho (-\tilde Q) \,,
\end{equation}
where the external field $h(\tilde Q)=h^*(-\tilde Q)$ couples to the
particle density $\rho (\tilde Q)=(\beta \nu )^{-1/2}\sum _{\tilde K,\sigma
} \psi ^*_\sigma (\tilde K)\psi _\sigma (\tilde K+\tilde Q)$. 
The density-density response function is determined by
\begin{equation}
\chi _{\rho \rho}(\tilde Q)= \langle \rho (\tilde Q) 
\rho (-\tilde Q) \rangle = 
{{\delta ^{(2)} \ln Z\lbrack h \rbrack } \over 
{\delta h^*(\tilde Q)\delta h(\tilde Q)}} \Biggl \vert _{h=0} \,,
\label{defcrr}
\end{equation}
where $Z\lbrack h\rbrack $ is the partition function in presence of the
source term. The RG process generates correction to the source field $h$
along with higher order terms in the source field. At step $t$, the total
action can be written as
\begin{equation}
S_{\Lambda (t)}-\sum _{\tilde Q} z_h(\tilde Q) h(\tilde Q) \rho (-\tilde Q)
-\sum _{\tilde Q} h^*(\tilde Q) h(\tilde Q) \chi (\tilde Q) +O(h^3) \,,
\end{equation}
where $S_{\Lambda (t)}$ is the action at step $t$ without the external
field $h$. $z_h(\tilde Q)$ and $\chi (\tilde Q)$ are both $t$-dependent
quantities. From (\ref{defcrr}), we deduce that $\chi _{\rho \rho}$ is the
FP value of $\chi$, i.e $\chi _{\rho \rho}=\chi ^*$. 
We consider the case where
only the Landau parameters $F^s_0$ and $F^a_0$ are non zero. $\Gamma
_{\sigma _i}$, $A$ and $B$ are then functions of $\tilde Q$ only and are
given by (\ref{AB00}).  

\begin{figure}
\epsfxsize9cm
\epsffile[53 242 559 550]{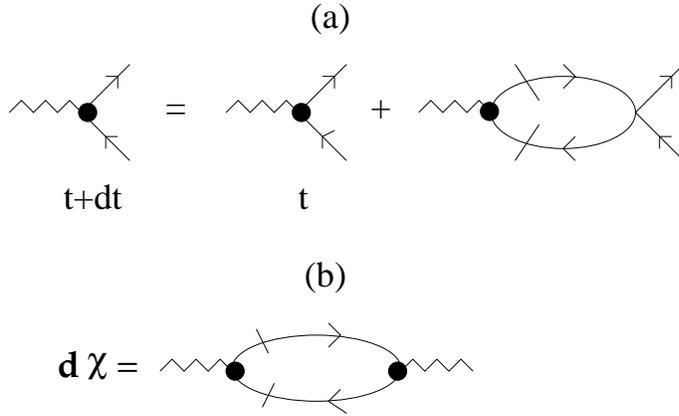}
\caption{(a) Diagrammatic representation of the renormalization of the
external field $h$ at one-loop order. The wavy line with the black dot
represents the renormalized external field $z_hh$. The slashed particle
lines indicate momenta in the infinitesimal shell $\Lambda (t)e^{-dt}\leq \vert
k\vert \leq \Lambda (t)$. (b) Diagrammatic representation of the
renormalization of $\chi $ at one-loop order.  }
\label{fig:response}
\end{figure}

The renormalization at one-loop order of $z_h$ (see Fig.\
\ref{fig:response}a) is given by
\begin{equation}
dz_h(\tilde Q)=z_h(\tilde Q) {T \over \nu } \sum _{\tilde K}' G(\tilde K)
G(\tilde K+\tilde Q) \sum _{\sigma '} \Gamma _{\sigma \sigma ',\sigma
'\sigma }(\tilde Q) \,,
\end{equation}
where $\sum '$ means that the sum is restricted to the degrees of freedom
which are in the infinitesimal momentum shell. As in
Sec.\ \ref{subsec:cm}, we impose on ${\bf K}$ to be in the infinitesimal shell
to be integrated out and let ${\bf K}+{\bf Q}$ free (and consider only the
case of small ${\bf Q}$). Using (\ref{bubbleZS}) and
$\sum _{\sigma '}\Gamma  _{\sigma \sigma ',\sigma '\sigma }(\tilde
Q)=A(\tilde Q)/N(0)$, we obtain
\begin{equation}
{{d\ln z_h(\tilde Q)} \over {dt}} = -{{\beta _R} \over {\cosh ^2(\beta _R)}}
\Omega _0(\eta ) A(\tilde Q). 
\end{equation}
As expected, the renormalization of $z_h$ involves only the charge
part ($A$) of the interaction. 
Because of the thermal factor $\beta _R/\cosh ^2(\beta _R)$ (which becomes
significantly different from zero only when $\Lambda (t) \lesssim T/v_F$), we
only need to know the expression of $A(\tilde Q)$ for $\Lambda (t)\lesssim
T/v_F$. Using $z_h(\tilde Q)\vert _{\tau =1}=1$ and (\ref{AB00}), 
we obtain
\begin{equation}
z_h(\tilde Q) = {1 \over {1+(1-\tau )F^s_0\Omega _0(\eta )}} \,.
\label{vertexpart}
\end{equation}
 
The generation of the term of order $O(h^2)$ is shown in Fig.\
\ref{fig:response}b. The RG equation for $\chi $ is given by
\begin{eqnarray}
d\chi (\tilde Q)&=&- {T \over \nu } \sum _{\tilde K,\sigma }'
G(\tilde K) G(\tilde K+\tilde Q) z^2_h(\tilde Q) \label{dchi} \\
&=& 2N(0)  {{\beta _R} \over {\cosh ^2(\beta _R)}} \Omega _0(\eta )
z^2_h(\tilde Q) dt \,.
\end{eqnarray}
Again the factor $\beta _R/\cosh ^2(\beta _R)$  allows us to use the
expression of $z_h(\tilde Q)$ for $\Lambda (t)\lesssim T/v_F$ (Eq.\
(\ref{vertexpart})). We obtain
\begin{equation}
\chi (\tilde Q)=2N(0) {{\Omega _0(\eta )(1-\tau )} \over 
{1+(1-\tau )F^s_0\Omega _0(\eta )}} \,,
\label{chi}
\end{equation}
using $\chi (\tilde Q)\vert _{\tau =1}=0$, which holds when $T\ll
v_F\Lambda _0$, since the external field $h(\tilde Q)$ couples only to
states which are within the thermal broadening of the Fermi surface. At the
FP ($\tau =0$), we recover the standard expression of the density-density
response function in the simple case we are considering ($F^{s,a}_l=0$ if
$l\neq 0$). From (\ref{chi}), we deduce the compressibility of the
Fermi liquid:
\begin{eqnarray}
\kappa ^*&=&\lim _{Q\to 0} \Bigl \lbrack \chi ^*(\tilde Q) \Bigl \vert 
_{\Omega =0} \Bigr \rbrack \nonumber \\
&=& {{2N(0)} \over {1+F^s_0}} \,, 
\end{eqnarray}
a result which holds whatever the values of the Landau parameters (i.e. 
$\kappa ^*$ is determined by $F_0^s$ only). 
The Pauli susceptibility can be obtained in a similar way by introducing an
external magnetic field which couples to the spin density. 

\subsection{Zero-temperature limit}

We discuss in this section the zero-temperature limit of the RG
equations. For simplicity, we only consider the contribution of the ZS graph
to $\Gamma ^Q$. This contribution can be written as
\begin{equation}
{{dA_l^Q} \over {dt}} \Biggl \vert _{\rm ZS}= 
-{{\beta _R} \over {\cosh ^2(\beta _R)}} {A_l^Q}^2  \,,
\label{EQztl}
\end{equation}
where $\beta _R=v_F\beta (t)\Lambda _0/2$ with $\Lambda _0$ the cut-off (which
is kept fixed through the rescaling procedure) and $\beta (t)=\beta e^{-t}$
the effective temperature at step $t$. Using $\lim _{\beta \to \infty }
\beta _R/\cosh ^2(\beta _R)=2\Lambda _0\delta (\Lambda _0)$, we obtain in
the zero temperature limit
\begin{equation}
{{dA_l^Q} \over {dt}} \Biggl \vert _{\rm ZS}= 
-2\Lambda _0\delta (\Lambda _0){A^Q_l}^2 =0 
\label{ZTL}
\end{equation}
for any finite value of the cut-off $\Lambda _0$. One can also obtain the 
preceding equation by taking the limit $T\to 0$ from the very beginning of
the calculation (i.e. in Eq.\ (\ref{action})). Eq.\ (\ref{ZTL}) disagrees with
the preceding sections where the limit $T\to 0$ was taken only at the end of
the calculation. The origin of this disagreement can be understood as
follows. The ZS graph describes processes where the two incoming particles
exchange a particle-hole pair at zero total momentum and energy. Assume that
the particle and the hole of this pair have momenta corresponding to the
same (band) energy $\epsilon $. Because of
the rescaling of the momenta, $\vert \epsilon \vert $ increases
($\epsilon '=e^{dt}\epsilon $ at each step of the renormalization). When it
reaches the value $\vert \epsilon \vert =\Lambda _0$, we obtain a finite
contribution to $dA^Q_l/dt\vert _{\rm ZS}$. At finite temperature, 
the particle and the hole of
the exchanged pair have energies within the thermal broadening of the Fermi
surface ($\vert \epsilon \vert \lesssim T$). However, at $T=0$, the
particle-hole pair has to lie exactly on the Fermi surface ($\epsilon
=0$). Under rescaling of the energies ($\epsilon '=e^{dt}\epsilon $),
$\epsilon =0$ remains unchanged and never reaches the value $\vert \epsilon 
\vert =\Lambda _0$. Hence the absence of flow for $A^Q_l$ in the ZS
channel. This unphysical result (first pointed out by Chitov and
S\'en\'echal \cite{Chitov95}) follows from the rescaling procedure keeping
the cut-off $\Lambda _0$ fixed which amounts to integrating all the degrees
of freedom except those on the Fermi surface. While this procedure is in
general perfectly valid, it fails at $T=0$ due to the somehow pathological
behavior of the ZS channel. A natural way
to avoid these difficulties is to use a finite temperature formalism taking
the limit $T\to 0$ only at the end of the calculations. \cite{Chitov95}
Alternatively, one
can determine $\Gamma (\theta _1,\theta _2;\tilde Q)$ for small but finite
$\tilde Q$ taking the limit $\tilde Q$ at the end of the calculation (a
finite $\bf Q$ ensures that the particle and the hole are not on the Fermi
surface. Under the rescaling procedure, their energies will therefore become
of the order of $\Lambda _0$). Another possibility would be not to follow the
rescaling procedure and to derive RG equations as a function of the effective
cut-off $\Lambda $. Eq.\ (\ref{EQztl}) is then replaced by
\begin{equation}
{{dA_l^Q} \over {d\Lambda }} \Biggl \vert _{\rm ZS} 
= {{v_F\beta /2} \over {\cosh ^2(v_F\beta
\Lambda /2)}} {A_l^Q}^2  \to _{(T\to 0)}
2\delta (\Lambda ){A_l^Q}^2 \,,
\end{equation}
which allows to integrate all the degrees of freedom since the cut-off can
reach the value $\Lambda =0$.

\section{A few remarks}
\label{sec:afr}
 
In this section, we discuss in detail some points which were only briefly
mentioned in Sec.\ \ref{sec:1L}.

\subsection{Content of the low-energy effective action} 

We first consider the problem which arises in the calculation of the ZS$'$ 
graph. When ${\bf Q}\rightarrow 0$, the ZS$'$ graph vanishes unless ${\bf K}_2
-{\bf K}_1 \rightarrow 0$ since both internal lines should have their momenta
in the infinitesimal shell near the cut-off. 
We therefore obtain a discontinuous contribution in 
the forward direction (${\bf K}_2={\bf K}_1$). For the same reason, when  
${\bf K}_2={\bf K}_1$, we obtain a discontinuity at ${\bf Q=0}$ when 
one varies ${\bf Q}$. The same problem arises in the calculation of the ZS 
graph considered as a function of ${\bf Q}$. These discontinuities are clearly
unphysical. 

Consider the one-loop diagrams of Fig.\ \ref{fig:corr_1loop}. All the internal 
momenta should be in the infinitesimal shell $\Lambda (t)e^{-dt}\leq 
\vert k\vert \leq
\Lambda (t)$ which has to be integrated out. We therefore consider only the
intermediate states where the particle and the hole (or both particles in the
case of the BCS graph) have the same energies (in absolute value,
i.e. $\vert \epsilon _1\vert =\vert \epsilon _2 \vert $). If, in the KW RG
method, we consider only these diagrams, we do not take into account processes
where the particle and the hole in the intermediate state do not have the
same energy. This reduction of the Hilbert space results in unphysical 
discontinuities. These discontinuities are suppressed if one includes in the
action three-body  interactions.\cite{Shankar94a} 

By iterating ``by
hand'' the RG equations, one can identify the diagrams which are effectively
considered via the RG approach. For instance, a one-loop RG calculation in
the ZS channel (ignoring the ZS$'$ and BCS channels) amounts to suming the
series of bubble diagrams in this channel. In the following we explicitly 
identify some of the diagrams generated by the RG equations to prove the
importance of three-body interactions. 

\begin{figure}
\epsfxsize14cm
\epsffile[-45 154 651 638]{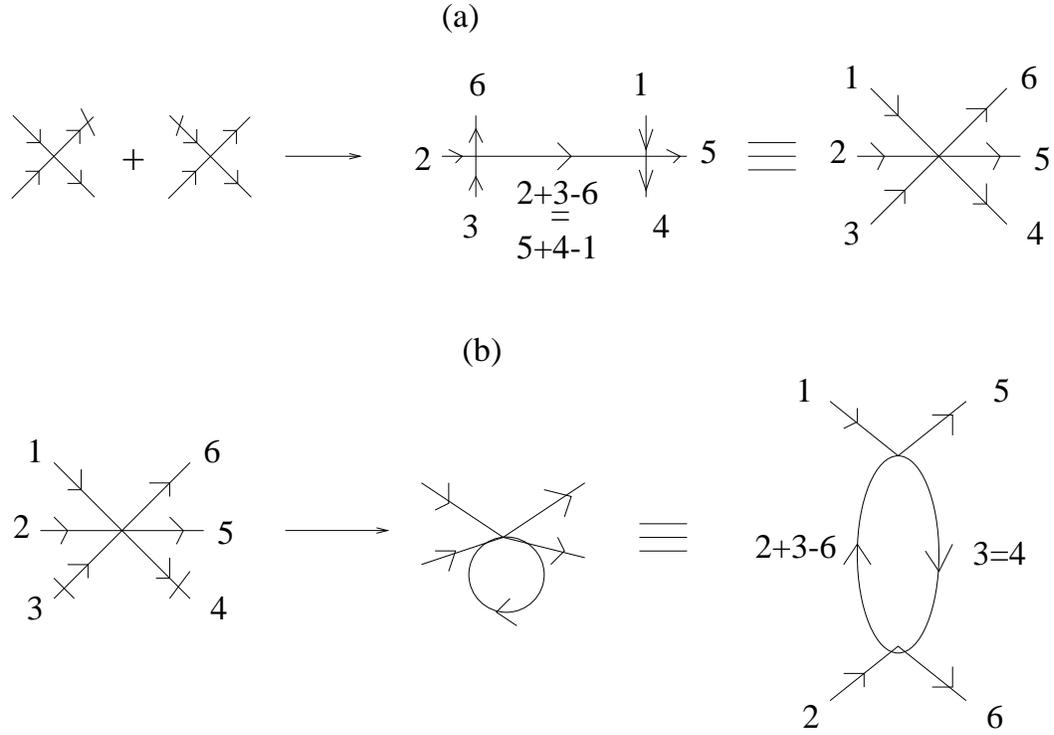}
\caption{(a) Generation of a six-leg diagram from four-leg diagrams. (b)
Generation of a four-leg diagram from a six-leg diagram. The slashed
particle lines indicate momenta in the infinitesimal shell $\Lambda
(t)e^{-dt}\leq \vert k\vert \leq \Lambda (t)$ }
\label{fig:3body}
\end{figure}

Consider the action $S=S_0+S_4+S_6$ where the quadratic and 
quartic parts, $S_0$ and $S_4$, are given by (\ref{action}) (where we now
note $U^{(4)}$ the coupling function of the two-body interaction). $S_6$ is
a three-body interaction given by
\begin{equation}
S_6= {1 \over {(3!)^2}}  {{T^2} \over {\nu ^2}} 
\sum _{\tilde K_1... \tilde K_6} U^{(6)}(\tilde K_i) 
\psi ^*(\tilde K_6) \psi ^*(\tilde K_5)\psi ^*(\tilde K_4)
\psi (\tilde K_3) \psi (\tilde K_2)\psi (\tilde K_1)
\delta _{1+2+3,4+5+6} \,,
\end{equation}
where we do not consider the spin dependence which is of no importance for our 
discussion. The function $\delta _{1+2+3,4+5+6}$ ensures the conservation of
momentum and energy and $U^{(6)}(\tilde K_i)\equiv U^{(6)}(\tilde K_1,...,
\tilde K_6)$. All wave-vectors satisfy $0\leq \vert k \vert \leq \Lambda
_0$. If we reduce the cut-off, $\Lambda _0'=\Lambda _0/s$ ($s>1$), and
rescale radial momenta, frequencies and fields in the usual way ($k'=sk$,
$\omega '=s\omega $, $\psi' =\psi $) to keep the quadratic action $S_0$ 
invariant, we obtain ${U^{(4)}}'=U^{(4)}$ and ${U^{(6)}}'=
U^{(6)}/s$. One usually concludes that $U^{(4)}$ is marginal and $U^{(6)}$ is
irrelevant, so that this latter can be neglected in the small $U^{(6)}$ 
limit. This conclusion is not correct if $U^{(6)}$ is not an analytic 
function of its arguments. This is precisely the situation we have to consider.
The RG generates a three-body interaction which is a singular function of its
arguments and turns out to be marginal. To lowest order, a three-body 
interaction is generated via the process shown in Fig.\ \ref{fig:3body}a. In 
this figure, the slashed lines indicate degrees of freedom in the infinitesimal
shell $\Lambda (t)e^{-dt}\leq \vert k\vert \leq \Lambda (t)$ which have 
to be integrated out. The other particle lines are all assumed to be below the
infinitesimal momentum shell ($\vert k \vert <\Lambda (t)
e^{-dt}$). The corresponding contribution to $S_6$ is of the type 
(ignoring sign and multiplicative factors)
\begin{eqnarray}
&& U^{(4)}(\tilde K_2,\tilde K_3,\tilde K_2+\tilde K_3-\tilde K_6,\tilde K_6;
t_{23\bar 6})
U^{(4)}(\tilde K_1,\tilde K_5+\tilde K_4-\tilde K_1,\tilde K_4,\tilde K_5;
t_{23\bar 6})   \nonumber \\ & \times & G(\tilde K_2+\tilde K_3-\tilde K_6)
\psi ^*(\tilde K_6) \psi ^*(\tilde K_5)\psi ^*(\tilde K_4)
\psi (\tilde K_3) \psi (\tilde K_2)\psi (\tilde K_1) \,,
\label{S6con}
\end{eqnarray}
where $t_{23\bar 6}$ is defined by $v_F\Lambda (t_{23\bar 6})=\vert \epsilon 
({\bf K}_2+{\bf K}_3-{\bf K}_6)\vert $. Because of the Green's function $G(
\tilde K_2
+\tilde K_3-\tilde K_6)=(i\omega _2+i\omega _3-i\omega _6-\epsilon ({\bf K}_2
+{\bf K}_3-{\bf K}_6))^{-1}$, this contribution is singular when $\epsilon ,
\omega \rightarrow 0$. In a dimensional analysis, $G(\tilde K_2+\tilde K_3-
\tilde K_6)$ yields an additional factor $s$ so that the contribution 
(\ref{S6con}) 
to $S_6$ is marginal and not irrelevant. Imagine that, once the contribution 
(\ref{S6con}) has been generated, one continues the renormalization process by
decreasing the cut-off below $\Lambda (t_{23\bar 6})$. If two (one incoming and
one outgoing) of the six external lines of the six-leg diagram of Fig.\ 
\ref{fig:3body}a have the same momentum and energy, for example $\tilde K_3=
\tilde K_4$, then for $v_F\Lambda (t)=\vert \epsilon ({\bf K}_3)\vert =\vert
\epsilon ({\bf K}_4)\vert $, this diagram generates a four-leg diagram
as shown in Fig.\ \ref{fig:3body}b, assuming that the other four external legs
are below the cut-off. (When the cut-off is between $\vert \epsilon ({\bf
K}_3)\vert  /v_F=\vert 
\epsilon ({\bf K}_4)\vert /v_F$ and $\Lambda (t_{23\bar 6})$, the contribution 
(\ref{S6con}) to $S_6$ does not renormalize.) 
We therefore obtain a contribution to $S_4$ of the type
\begin{eqnarray}
&& U^{(4)}(\tilde K_2,\tilde K_3,\tilde K_2+\tilde K_3-\tilde K_6,\tilde K_6;
t_{23\bar 6})
U^{(4)}(\tilde K_1,\tilde K_5+\tilde K_4-\tilde K_1,\tilde K_4,\tilde K_5;
t_{23\bar 6})   \nonumber \\ & \times & G(\tilde K_2+\tilde K_3-\tilde K_6)
G(\tilde K_3) 
\psi ^*(\tilde K_6) \psi ^*(\tilde K_5)
 \psi (\tilde K_2)\psi (\tilde K_1) \,.
\end{eqnarray}
The important point is that $\vert \epsilon ({\bf K}_3)\vert \neq \vert 
\epsilon ({\bf K}_2+{\bf K}_3-{\bf K}_6)\vert $. Thus, via the three-body
interaction, we have generated the ``missing'' processes where the particle
and the hole in the intermediate state do not have the same energy. This
shows that it is  necessary to
consider three-body interactions to generate all the one-loop diagrams in the
KW RG approach. The calculation of one-loop diagrams 
generated from six-leg vertices would be in practice very difficult. The reason
is that, in the above example (Fig.\ \ref{fig:3body}), the 
contribution to $U^{(4)}$ at $\Lambda (t)=\vert \epsilon ({\bf K}_3)\vert =
\vert \epsilon ({\bf K}_4)\vert $ involves 
$U^{(4)}(t_{23\bar 6})$ where $t_{23\bar 6}<t$ (i.e. $\Lambda
(t_{23\bar 6})>\Lambda (t)$). Therefore, $dU^{(4)}(t)/dt$ is not a function of 
$t$ only but depends also on $t'<t$: $U^{(4)}(t)$ is determined by an 
integro-differential equation. Nevertheless, if $\vert \epsilon ({\bf
K}_3)\vert \simeq \vert \epsilon ({\bf K}_2+{\bf K}_3-{\bf K}_6)\vert $, we
have $\Lambda (t_{23\bar 6})\simeq \Lambda (t)$ and we can make the
approximation $U^{(4)}(t_{23\bar 6})=U^{(4)}(t)$. $dU^{(4)}(t)/dt$ is then
entirely determined by $U^{(4)}(t)$. In other words, we have approximated
the integro-differential equation which determines $U^{(4)}(t)$ by a
differential equation. It is clear that this approximation amounts to
calculating $dU^{(4)}(t)/dt$ directly from the one-loop diagram $\propto
{U^{(4)}}^2$ (i.e. without considering $U^{(6)}$) imposing ${\bf K}_3$ to
be at the cut-off and relaxing the analog constraint for ${\bf K}_2+{\bf
K}_3-{\bf K}_6$. This is precisely what we have done in Sec.\
\ref{sec:1L} to calculate the ZS$'$ graph for ${\bf K}_1^F-{\bf K}_2^F\neq
0$ or the ZS graph at finite $\tilde Q$. For this approximation to be
meaningful, one should nonetheless verify that the same differential
equation is obtained if, for example, one chooses to put one particle (${\bf
K}_3$) slightly below the cut-off ant the other one (${\bf K}_2+{\bf
K}_3-{\bf K}_6$) slightly above. While such a problem does not arise in the
case we consider in this paper, it has been shown by Hubert that in general
different choices may lead to different results. \cite{Hubert}

\begin{figure}
\epsfxsize8cm
\epsffile[8 88 604 704]{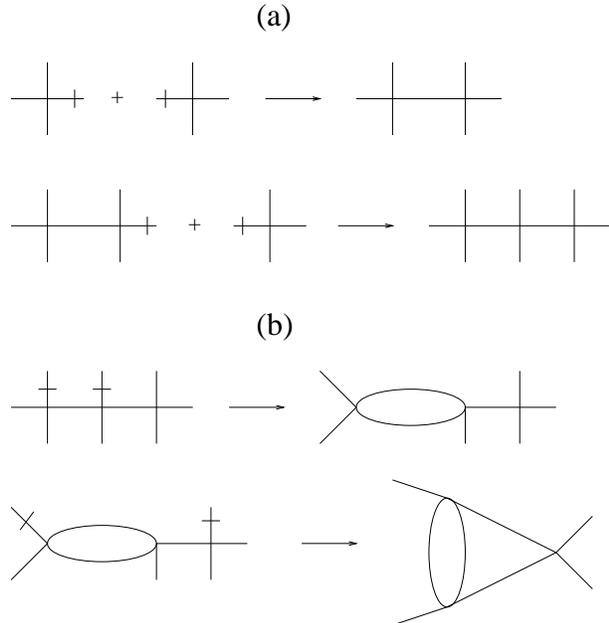}
\caption{(a) generation of an eight-leg diagram from four-leg diagrams. (b)
generation of a four-leg diagram from an eight-leg diagram. }
\label{fig:4body}
\end{figure}

In the same way, the RG generates eight-leg vertices which, for the reason 
discussed above, are marginal. These vertices generate in turn four-leg
vertices. Fig.\ \ref{fig:4body} shows how a
two-loop diagram is generated in this way. This two-loop diagram cannot be
generated  directly (i.e. from the integration of only one infinitesimal
momentum shell) because of the constraints
imposed by momentum  conservation (in the diagram of Fig.\ \ref{fig:4body}b, it
is not possible to have  all the internal momenta in the shell to be
integrated out).  

$n$-body ($n>2$) interactions are also generated by the integration of
high-energy degrees of freedom $\vert \epsilon \vert >E_0$ ($E_0=v_F\Lambda 
_0$ assuming a circular Fermi surface) which is the necessary step to obtain
the low-energy effective action. We note $S_{\rm micro}(\psi ^*,\psi )$ the
exact microscopic action. $\psi ^{(*)}\equiv \psi ^{(*)}({\bf K},\omega )$ 
where ${\bf K}$ belongs to the first Brillouin zone (assuming for simplicity a
single band). If we note $\psi ^{(*)}_>$ ($\psi ^{(*)}_<$) the fields with
$\vert \epsilon ({\bf K})\vert >E_0$ ($\vert \epsilon ({\bf K})\vert <E_0$),
then the low-energy effective action is defined by
\begin{equation}
e^{-S(\psi ^*_<,\psi _<)}= \int {\cal D}\psi ^*_> {\cal D}\psi _> 
e^{-S_{\rm micro}(\psi ^*_>,\psi _>;\psi ^*_<,\psi _<)} \,.
\label{Seff}
\end{equation}
This partial integration generates vertices at all order for the $\psi
^{(*)}_<$'s even if the microscopic action $S_{\rm micro}$ contains only
a two-body interaction.\cite{Popov87} As discussed above, some of these 
$n$-body interactions are marginal due to a non trivial dependence on the 
external variables. They should therefore be retained in the low-energy 
effective action. Physically, the role of these $n$-body interactions is
clear. For example, at one-loop order, the three-body interaction is necessary
to take into account processes where, in the intermediate state, the
particle is below the initial cut-off ($\vert \epsilon \vert <E_0$) while
the hole (or the other  particle) is above the initial cut-off ($\vert
\epsilon \vert >E_0$).

\subsection{Self-energy corrections}

Until now, we have not considered the renormalization of the one-particle
propagator. The corresponding diagram at one and two-loop orders are shown in
Fig.\ \ref{fig:self}. The two-loop diagram vanishes because it is not
possible to have all internal momenta in the shell to be integrated
out. This is also true for higher order diagrams. The only diagram which
does not vanish is the one-loop diagram. In general, non-trivial
self-energy corrections (finite life-time and wave-function renormalization)
originate in the dependence of $U^{(4)}$ on $\tilde
Q$. The integration of high-energy degrees of freedom also
generates a wave-function renormalization. However, it cannot induce a
finite life-time for states near the Fermi surface, since this effect comes
from real transitions occuring at low-energy. Since the dependence of
$U^{(4)}$ on $\tilde Q$ arises through the consideration of three-,
four-... $n$-body interactions, the inclusion of $U^{(6)}$, $U^{(8)}$... in
the action is also very important for the calculation of the self-energy.   

The role of $n$-body ($n>2$) is then crucial in the KW RG 
approach. They should be taken into account according to the order to which
the calculation is to be done. For example, for a two-loop order calculation, 
one should include in the action four-body interactions. In practice, the
KW RG method, as described here, would be very difficult to apply. We have
shown in Sec.\ \ref{sec:1L} how a one-loop order RG calculation can be
(approximately) done without considering three-body interactions (see also
Ref.\ \onlinecite{Bourbon91}). Moreover, Bourbonnais and Caron have 
shown how the KW RG approach 
can be modified to allow a two-loop calculation. \cite{Bourbon91} Their
method will be used in Sec.\ \ref{sec:qpp} to obtain the quasi-particle
life-time and the wave-function renormalization factor.  

\subsection{Interference between channels}
\label{subsec:ibc}

In the KW RG approach, the renormalization of the forward scattering
coupling function involves mainly the ZS channel. As discussed above, the
interference with the ZS$'$ channel requires the consideration of (at least) 
three-body interactions. This holds also for the interference between the ZS
and BCS channels. This means that the interference between channels
involves not only intermediate states at a given energy ($\Lambda (t)$) but
also intermediate states at higher energy ($\vert \epsilon \vert >\Lambda
(t)$). In other words, the interference between channels is
``frustrated''. This situation is characteristic of a two-dimensional
system with a circular 
Fermi surface and is at the basis of the validity of FLT. It is also
what justifies the use of a one-loop RG approach in the BCS channel only 
\cite{Shankar94} (i.e. a ladder diagrams summation (or RPA approximation) in
the conventional diagrammatic language) to study the BCS instability. For
more complicated Fermi surfaces, the 
interference between channels may become important. An example would be a
two-dimensional conductor near half-filling where $d$-wave superconductivity
can be induced by the exchange of spin fluctuations. (This Kohn-Luttinger
effect \cite{Kohn65,Shankar94} always exists but is expected to lead to
extremely small critical temperature in the case of a circular Fermi surface.) 
Another example is given by one-dimensional conductors where the different
channels of correlation strongly interfere forbidding any RPA-like
calculations. \cite{Solyom79,Bourbon91}

The consideration of the ZS$'$ channel in Sec.\ \ref{sec:1L} was motivated
by symmetry considerations. For $\vert \theta _1-\theta _2\vert \ll T/E_F$ 
($\theta _1-\theta _2$ being the
angle between the two incoming particles), the interference between the ZS
and ZS$'$ channel is not frustrated. This induces a particular behavior of the
two-particle vertex function around $\theta _1-\theta _2=0$ and ensures that
$\Gamma ^Q$ satisfies the Pauli principle. 

It turns out that the frustration of the interference between the ZS and BCS
channels also disappears for $\theta _1-\theta _2 \sim \pi $. For these
values of $\theta _1-\theta _2$, $\Gamma
(\theta _1,\theta _2;\tilde Q)$ can be seen both as a forward scattering
coupling function or as a BCS coupling function (Shankar's $V$ function
\cite{Shankar94}). We therefore expect a particular behavior of $\Gamma
(\theta _1-\theta _2\sim \pi )$. Such a behavior has been found for a dilute
Fermi gas 
(where a well-controlled low-density expansion can be made) in both
three-dimensional \cite{Abrikosov58,Lifshitz80}
and two-dimensional systems. \cite{Engelbrecht92}

\subsection{Field theory approach}

We briefly discuss in this section the differences between the KW approach
(which is used in the rest of this paper) and the FT
approach. In the latter, one calculates $n$-point vertices in a cut-off
theory as a function of the bare couplings. By requiring the renormalized
vertices to be independent of the cut-off, one obtains the evolution of the
bare vertices with the cut-off. \cite{Lebellac91,Shankar94} 
Consider for example the
renormalization of $\Gamma ^Q$. At one-loop order, the renormalized
two-particle vertex function can be written as
\begin{equation}
\Gamma ^Q\Bigl \vert _R = \Gamma ^Q +\delta \Gamma ^Q\Bigl \vert _{{\rm
ZS,ZS}',{\rm BCS}}  \,,
\end{equation}
where $\delta \Gamma ^Q$ is the correction calculated with a cut-off
$\Lambda (t)=\Lambda _0e^{-t}$. The dependence of $\Gamma ^Q$ on $t$ is then
obtained from the equation:
\begin{equation}
{d \over {dt}} \Gamma ^Q\Bigl \vert _R =
{d \over {dt}} \left ( \Gamma ^Q +\delta \Gamma ^Q\Bigl \vert _{{\rm
ZS,ZS}',{\rm BCS}} \right )=0\,.  
\end{equation}
Consider now the ZS$'$ graph where the particle and the hole in the
intermediate state have energies $\epsilon _1$ and $\epsilon _2$ with
$\vert \epsilon _1\vert \neq \vert \epsilon _2\vert $. In the FT approach,
this graph contributes to the RG flow when $\max (\vert \epsilon _1\vert
,\vert \epsilon _2\vert )=\Lambda (t)$. Thus, as pointed out by Shankar,
\cite{Shankar94} $d\Gamma ^Q/dt$ is given by the sum of all graphs where one
momentum is at the cut-off while the others are below. While the
KW and FT approaches are clearly equivalent for the ZS graph for ${\bf Q}\to
0$ (since if one
momentum of the particle-hole bubble is at the cut-off, then momentum
conservation ensures that the other one is also at the cut-off), they in
general differ. In the FT approach, it is clear that 
there is no need to consider three-, four- ... body interactions
contrary to the KW approach. This also means that $\Gamma
^Q(t)$ does not contain the same information in the KW and FT approaches. In
particular, the low-energy effective action (\ref{action}) (defined for a
given cut-off $\Lambda _0$) is not the same in both approaches. The
two-body interaction $U_{\sigma _i}$ is different and the action contains
$n$-body interactions ($n>2$) in the KW approach. Notice that when one
defines the low-energy effective action by (\ref{Seff}) (i.e. by integrating
out (in a functional sense) the high-energy degrees of freedom), one always
uses (implicitly) the KW approach. 

An interesting aspect of the FT approach is that the frustration of the
interference between
channels appears very naturally since it always involves the small parameter
$\Lambda /K_F$ because of phase space restriction.\cite{Shankar94} (Notice that
the small parameter $\Lambda /K_F$ never appears in the KW approach.) 
For instance,
the contribution of the ZS$'$ graph to the renormalization of $\Gamma ^Q$ is
of order $\Lambda /K_F$ with respect to the one of the ZS graph. We pointed
out in Sec.\ \ref{subsec:ibc} that the interference between channels in a
two-dimensional system with a circular Fermi surface is mainly determined by
high-energy states. In the KW approach, its description requires the
consideration of (at least) three-body interactions. In the FT approach,
processes involving both low and high-energy states are integrated out in the
early stage of the renormalization procedure. The interference left at
low-energy is suppressed by the small parameter $\Lambda  /K_F$. As
discussed by several authors, \cite{Shankar94,Feldman95} 
this latter property can be used to control
the perturbation expansion in a way similar to the $1/N$ expansion in
statistical mechanics.

\section{Beyond one-loop}
\label{sec:bol}

In Sec.\ \ref{sec:1L}, we used the singularity arising at low temperature 
in the flow of
$\Gamma _{\sigma _i}(\theta _1,\theta _2;\tilde Q)$ for $\Lambda
(t)\rightarrow 0$ to recover at one-loop 
order the results of the microscopic FLT. In this section, we show that
we can solve in the same way the RG equations at all orders if we assume
the existence of well defined quasi-particles (near the Fermi surface) and 
that the only singular contribution to the RG flow is due to the one-loop ZS
graph (as shown in Sec.\ \ref{sec:1L}, the contribution of the ZS$'$
graph to the flow of $\Gamma _{\sigma _i}(\theta _1,\theta _2;\tilde Q)$ can be
considered as regular if we are interested in quantities which involve all
the values of $\theta _1-\theta _2$). Let us
stress again that our aim is not to calculate FP quantities as a function of
the bare parameters of the action, but to relate physical quantities with
${\Gamma ^\Omega _{\sigma _i}}^*$. In the following, the action is not
restricted to a two-body interaction (as in (\ref{action})), but contains
also three-, four-, ...$n$-body interactions which are generated either by
the RG process or the integration of high-energy degrees of freedom ($\vert
\epsilon \vert >v_F\Lambda _0$) as discussed in Sec.\ \ref{sec:afr}. 

We first consider the renormalization of the one-particle Green's
function $G$. We note $z(t)$ the quasi-particle renormalization factor (and
assume $z(t)>0$) and write the Green's
function as $G(\tilde K)=(i\omega -v_F(t)k)^{-1}$ where the Fermi velocity
$v_F(t)$ depends on the flow parameter $t$. This form, which assumes a
proper rescaling of the fields, will be justified
below. The integration of the degrees of freedom in the infinitesimal
momentum shell modifies the Green's function:
\begin{equation}
G(\tilde K)^{-1} \Bigr \vert _{t+dt}=i\omega -v_F(t)k-d\Sigma (k\omega ) \,,
\end{equation}
where the self-energy correction $d\Sigma (k\omega )$ depends only on $k$
because of rotational invariance. We analyze the self-energy following Ref.\
\onlinecite{Shankar94}. We Taylor expand $d\Sigma (k\omega )$ as follows:
\begin{equation}
d\Sigma (k\omega )=d\Sigma (00)+i\omega {{\partial d\Sigma (k\omega )} \over
{\partial i\omega }} \Biggr \vert _{i\omega =k=0} +k {{\partial d\Sigma
(k\omega )} \over {\partial k}} \Biggr \vert _{i\omega =k=0} + \cdot \cdot
\cdot \,,
\end{equation}
where the dots denote irrelevant terms at tree-level. 
In the following, we neglect any effect
associated with a finite life-time of the quasi-particles, i.e. we assume
that $d\Sigma (00)$, $\partial d\Sigma (k\omega )/\partial i\omega \vert _{i
\omega =k=0}$, and $\partial d\Sigma (k\omega )/\partial k\vert _{i\omega
=k=0}$ are real. This is justified when the scattering rate is
much smaller than $\omega \sim T$. In a two-dimensional Fermi liquid, this 
latter is known to be of order $T^2\ln T$ (see Sec.\ \ref{sec:qpp}) and can
be neglected at low temperature. Ignoring irrelevant
terms and $d\Sigma (00)$, which corresponds to a non essential shift of the
chemical potential, we write the Green's function as
\begin{equation}
G(\tilde K)^{-1} \Bigr \vert _{t+dt}=i\omega z^{-1}(dt)-v_F(t)kz_m^{-1}(dt)
\,,
\end{equation}
where 
\begin{eqnarray}
z^{-1}(dt) &=& 1-{{\partial d\Sigma (k\omega )} \over
{\partial i\omega }} \Biggr \vert _{i\omega =k=0} \,, \nonumber \\ 
z_m^{-1}(dt) &=& 1+ {1 \over v_F(t)} {{\partial d\Sigma
(k\omega )} \over {\partial k}} \Biggr \vert _{i\omega =k=0} \nonumber \\
&=& 1+{{m(t)} \over {K_F}}  {{\partial d\Sigma
(k\omega )} \over {\partial k}} \Biggr \vert _{i\omega =k=0} \,.
\label{RGEQ101}
\end{eqnarray}
We have introduced the effective mass $m(t)=K_F/v_F(t)$. Since the
coefficients of $i\omega $ and $k$ are modified by different parameters, no
rescaling will keep the quadratic part of the action invariant. If one
chooses to rescale the fields to keep the coefficient of $i\omega $ fixed at
unity, i.e. ${\psi ^{(*)}}'=\lbrack z(dt)\rbrack ^{-{1 \over 2}}{\psi
^{(*)}}$, we obtain
\begin{equation}
G(\tilde K)^{-1} \Bigr \vert _{t+dt}=i\omega-v_F(t)kz_m^{-1}(dt)z(dt) \,.
\label{Gdt}
\end{equation}
The preceding equation shows that the quasi-particle form of the
one-particle propagator, $G(\tilde K)=(i\omega -v_F(t)k)^{-1}$, is conserved
if one ignores irrelevant terms and finite life-time effects. The condition
$z(t)>0$ then ensures the existence of well-defined quasi-particles. 
The rescaling of the fields modify the wave-function renormalization factor
which becomes 
\begin{equation}
z(t+dt)=z(dt)z(t) \,. 
\label{RGEQ11}
\end{equation}
Moreover, from (\ref{Gdt}) one obtains the following RG equation for the
Fermi velocity:
\begin{equation}
v_F(t+dt)=z(dt)z_m^{-1}(dt)v_F(t) \,,
\label{RGEQ12}
\end{equation}
or, equivalently,
\begin{equation}
m(t+dt)= z^{-1}(dt)z_m(dt)m(t) \,.
\end{equation}

We now consider the renormalization of the two-particle vertex function 
$\Gamma ^Q$ (the renormalization of $\Gamma (\theta _1,\theta
_2;\tilde Q)$ is discussed below). We make the assumption that
all graphs, except the one-loop ZS graph, are well-behaved for $\tilde Q\to
0$ and give a smooth contribution (with respect to $\Lambda (t)$) to the RG
flow of $\Gamma _{\sigma _i}(\theta _1,\theta _2;\tilde Q)$. Note that the
same kind of assumption is made in the microscopic FLT. The RG flows
of $\Gamma ^Q$ and $\Gamma ^\Omega $ are then determined by (before the
rescaling of the fields) 
\begin{eqnarray}
{{d\Gamma ^Q_{\sigma _i}} \over {dt}} &=& 
{{d\Gamma ^Q_{\sigma _i}} \over {dt}} \Biggl \vert _{{\rm 
ZS}} +
{{d\Gamma ^Q_{\sigma _i}} \over {dt}} \Biggl \vert 
_{{\rm ZS}',{\rm BCS},2\, {\rm loops}...} \label{RGEQ14} \\ 
{{d\Gamma ^\Omega _{\sigma _i}} \over {dt}} &=& 
{{d\Gamma ^\Omega _{\sigma _i}} \over {dt}} \Biggl \vert 
_{{\rm ZS}',{\rm BCS},2\, {\rm loops}...}
= {{d\Gamma ^Q_{\sigma _i}} \over {dt}} \Biggl \vert 
_{{\rm ZS}',{\rm BCS},2\, {\rm loops}...}  
\label{RGEQ15}
\end{eqnarray} 
The contribution of the one-loop ZS graph (first term of the rhs of
(\ref{RGEQ14})), which gives the only singular contribution to the flow of
$\Gamma ^Q$, has been separated from the non-singular contributions. 
Eqs.\ (\ref{RGEQ14},\ref{RGEQ15}) can be combined to obtain 
\begin{equation}
{{d\Gamma ^Q_{\sigma _i}} \over {dt}}=
{{d\Gamma ^\Omega _{\sigma _i}} \over {dt}}+
{{d\Gamma ^Q_{\sigma _i}} \over {dt}} \Biggl \vert _{\rm ZS}\,.
\label{RGEQ16}
\end{equation}
According to our assumption, $\Gamma ^\Omega _{\sigma _i}$ is a non-singular
function of $\Lambda (t)$ since it does not receive any contribution from
the one-loop ZS graph (see Sec.\ \ref{subsec:GQGO}). 
Using the results of Sec.\ \ref{subsec:GQGO}, we have 
\begin{equation}
{{d\Gamma ^Q_{\sigma _i}(l)} \over {dt}} \Biggl \vert _{\rm 
ZS} = - {{N(0)\beta _R} \over {\cosh ^2(\beta _R)}} \sum _{\sigma ,\sigma
'} \Gamma ^Q_{\sigma _1\sigma ',\sigma \sigma _4}(l) \Gamma ^Q_{\sigma \sigma
_2,\sigma _3\sigma '}(l) \,,
\label{RGEQ17}
\end{equation}
where $N(0)=K_F/2\pi v_F(t)$ and $\beta _R=v_F(t)\beta \Lambda (t)/2$ since the
one-particle Green's function has the form $G(\tilde K)=(i\omega
-v_F(t)k)^{-1}$. Eq.\ (\ref{RGEQ16}) can be written in the compact form
\begin{equation}
\Gamma ^Q_{\sigma _i}(l,t+dt)=z_{\sigma _i}^{(\Gamma )}(l,dt) 
\Gamma ^Q_{\sigma _i}(l,t) \,. 
\end{equation}
After the rescaling of the fields, ${\psi ^{(*)}}'=\lbrack z(dt)\rbrack ^{-{1
\over 2}}{\psi ^{(*)}}$, we obtain the RG equation
\begin{equation}
\Gamma ^Q_{\sigma _i}(l,t+dt)=z(dt)^2 z_{\sigma _i}^{(\Gamma )}(l,dt) 
\Gamma ^Q_{\sigma _i}(l,t) \,. 
\label{RGEQ18}
\end{equation}

The RG equations at all orders are given by
(\ref{RGEQ11},\ref{RGEQ12},\ref{RGEQ18}) which (together with the assumption
that $\Gamma ^\Omega $ is a smooth function of $\Lambda (t)$) constitute the
basis of FLT in the RG language. $z(t)$ and $v_F(t)$ are determined
by the dependence of $\Gamma $ on $\tilde Q$ through the self-energy
$d\Sigma (k\omega )$. Since the singularity in the flow of $\Gamma $ is
weakened at finite $\tilde Q$ (only the flow of $\Gamma ^Q$ presents a
singularity $\sim \delta (\Lambda )$ for $T\to 0$), $z(t)$ and $v_F(t)$ are
smooth functions of the cut-off $\Lambda (t)$. At low temperature and for
$\Lambda (t)\lesssim T/v_F$, they can therefore be approximated by their FP
values: $z(t)\vert _{\Lambda (t)\sim T/v_F}\simeq z^*$ and $v_F(t)\vert
_{\Lambda (t)\sim T/v_F}\simeq v_F^*$. Consequently, we have
$z(dt)=z_m(dt)\simeq 1$ for $\Lambda (t)\lesssim T/v_F$. Since the contribution
of the ZS graph becomes significantly different from zero only when $\Lambda
(t)\lesssim T/v_F$, Eqs.\ (\ref{RGEQ11},\ref{RGEQ12},\ref{RGEQ18}) reduce to
(\ref{RGEQ16}) with $z(t)$ and $v_F(t)$ equal to their FP values. Eq.\ 
(\ref{RGEQ16}) is similar to (\ref{RGEQ1}) and can be solved in the
same way which leads again to (\ref{PV1}). The expression of ${\Gamma ^Q}^*$ as
a function of the Landau parameters is recovered if these latter are defined
by $f_{\sigma _i}(\theta )={\Gamma ^\Omega _{\sigma _i}}^*(\theta )$. 
Since the $\psi ^{(*)}$'s have been rescaled at each step of the
renormalization, we eventually come to
\begin{equation}
f_{\sigma _i}(\theta )={z^*}^2{\Gamma _{\sigma _i}^\Omega }^*(\theta )\,, 
\label{FRG}
\end{equation}
where ${\Gamma _{\sigma _i}^\Omega }^*(\theta )$ now refers to the bare
fermions. Eq.\ (\ref{FRG}) defines the Landau function in the RG approach. 

The renormalization of $\Gamma _{\sigma _i}(\theta _1,\theta _2;\tilde Q)$
can be discussed in the same way. However, for finite $Q$, the
resolution of the RG equations becomes less accurate because the singularity
$(\beta /4)\cosh ^{-2}(\beta v_Fk/2)\vert _{T\to 0}=\delta (v_Fk)$ is weakened
(see Sec.\
\ref{subsec:cm}). In particular, since $z(t)$ and $v_F(t)$ depend on $\Gamma
_{\sigma _i}(\theta _1,\theta _2;\tilde Q)$ for finite $\tilde Q$, the
replacement $z(t)\to z^*$ and $v_F(t)\to v_F^*$ in the RG equation for
$\Gamma _{\sigma _i}(\theta _1,\theta _2;\tilde Q)$ is not exact any more.
Eqs.\ (\ref{RGEQ11},\ref{RGEQ12},\ref{RGEQ18}) should be
considered together if they were to be solved exactly. This would lead to
complicated coupled equations for $\Gamma _{\sigma _i}(\theta _1,\theta
_2;\tilde Q)$, $z(t)$ and $v_F(t)$. 

Thus, the relations between ${\Gamma ^\Omega }^*$ and physical quantities
obtained at one-loop order (Sec.\ \ref{sec:1L}) are essentially unchanged by
higher order contributions although of course the expression of ${\Gamma
^\Omega }^*$ as a function of the microscopic parameters is changed. (The
only change is that the ``bare'' quantities $v_F$ and $z_{\Lambda _0}$ are
replaced by their FP values $v_F^*$ and $z^*$.) This is
a direct consequence of our assumption according to which the only singular
contribution to the RG flow comes from the one-loop ZS graph.

\section{Quasi-particle properties} 
\label{sec:qpp}

In this section, we calculate the quasi-particle life-time and
renormalization factor. As discussed in Sec.\ \ref{sec:afr}, the self-energy
is obtained from the one-loop diagram (since all higher order diagrams
vanish). Thus, one possibility to obtain the self-energy would be to
introduce the expression of $\Gamma _{\sigma _i}(\theta _1,\theta _2;\tilde
Q)$ obtained in Sec.\ \ref{subsec:cm} in the one-loop diagram. In practice,
such a procedure turns out to be very difficult. We follow in this section
an alternative method introduced by Bourbonnais and Caron. \cite{Bourbon91}
These authors have shown how the KW RG approach can be
modified in order to easily obtain the two-loop corrections and therefore
the self-energy at this order. The main idea is to make a distinction between
band momenta and transfer momenta. At each step of the renormalization, one
integrates the fermion field degrees of freedom corresponding to a band
momentum in the infinitesimal shell $\Lambda (t)e^{-dt}\leq \vert k\vert \leq
\Lambda (t)$ to be integrated out. No cut-off is imposed on the transfer
momenta which are let free. In principle, it is necessary to impose 
an additional constraint on
the transfer momenta in order to ensure that every degree of freedom is
integrated once and only once (at a given order). In practice, one can ignore
this constraint which is automatically taken into account through the Fermi
factors. The application of this method to one-dimensional systems has
been very successful since it has allowed to recover the results of the
multiplicative renormalization group (MRG) approach. \cite{Solyom79}

\begin{figure}
\epsfxsize8cm
\epsffile[26 282 587 510]{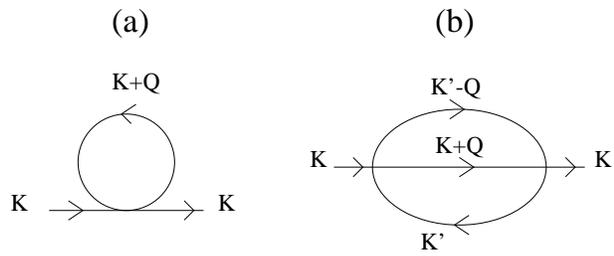}
\caption{One-loop (a) and two-loop (b) diagrams for the self-energy. }
\label{fig:self}
\end{figure}

We consider the simple case where only
$F^s_0$ and $F^a_0$ are non zero. As discussed at the end of Sec.\
\ref{sec:bol}, the calculation of quantities involving $\Gamma (\theta
_1,\theta _2;\tilde Q)$ with finite $Q$ requires to consider on the same
footing the renormalization of $z(t)$ and $v_F(t)$. For simplicity, we
assume that all quantities can be calculated with the FP form of the
one-particle propagator, $G(\tilde K)=(i\omega -v_F^*k)^{-1}$, ignoring any
finite life-time as can be justified at low temperature.  
Moreover, we do not take into account the effect of the
rescaling of the fields, ${\psi ^{(*)}}'=\lbrack z(dt)\rbrack ^{-{1 \over 2}}
\psi ^{(*)}$, on the RG equation of $\Gamma (\theta _1,\theta _2;\tilde
Q)$. These approximations are expected to be correct as long as one
considers only small transfer momentum ${\bf Q}$ in the diagram for the
self-energy (Fig.\ \ref{fig:self}b). This can be achieved by introducing a
cut-off $Q_c$ for the transfer momenta. We can then use the results of
Secs.\ \ref{subsec:cm} and \ref{subsec:ddrf} which hold in the limit of
small $Q$. The cut-off $Q_c$ is also introduced in the microscopic FLT
where the self-energy is usually obtained by dressing the particle line with
a charge or spin fluctuation propagator with $Q<Q_c$. \cite{Stamp92}

The self-energy correction is  given by (Fig.\ \ref{fig:self}b)
\begin{eqnarray}
d\Sigma (k\omega )&=&-{{T^2} \over {2\nu ^2}} \sum _{\tilde Q} \sum
_{\tilde K'}' \sum _{\sigma _1\sigma _2\sigma _3} G(\tilde K+\tilde
Q)G(\tilde K')G(\tilde K'-\tilde Q) \Gamma _{\sigma \sigma _1,\sigma
_2\sigma _3}(\tilde Q) \Gamma _{\sigma _3\sigma _2,\sigma
_1\sigma }(\tilde Q)  \nonumber \\
&=& -{1 \over {4N^2(0)}} {{T^2} \over {\nu ^2}} \sum _{\tilde Q} \sum
_{\tilde K'}'  G(\tilde K+\tilde Q)G(\tilde K')G(\tilde K'-\tilde Q) 
\Bigl (A^2(\tilde Q)+3B^2(\tilde Q)\Bigr ) \,,
\end{eqnarray}
where $\Gamma (\tilde Q)$, $A(\tilde Q)$ and $B(\tilde Q)$ are given by 
(\ref{AB00}). In the above equation, the sum over the transfer momentum ${\bf
Q}$ is free (with $Q<Q_c$) while the sum over the band momentum ${\bf K}'$ is
restricted to $\Lambda (t)e^{-dt}\leq \vert k'\vert \leq \Lambda (t)$. In the
following we consider only the charge part of the interaction (i.e. we put
$B=0$). $d\Sigma (k\omega )$ can be expressed as a function of the
density-density response function obtained in Sec.\ \ref{subsec:ddrf}. Using
(\ref{AB00},\ref{vertexpart},\ref{dchi}), we have 
\begin{equation}
d\chi (\tilde Q)=-2 {T \over \nu }\sum _{\tilde K'}' G(\tilde K')G(\tilde
K'+\tilde Q){{A^2(\tilde Q)} \over {{F^s_0}^2}} \,.
\end{equation}
We therefore obtain
\begin{equation}
d\Sigma (k\omega )= {T \over {2\nu }} \sum _{\tilde Q} G(\tilde
K+\tilde Q) {f_0^s}^2 d\chi (\tilde Q) \,,
\end{equation}
where $f_0^s=F_0^s/2N(0)$. Integrating this equation, we obtain the FP value
of the self-energy:
\begin{equation}
\Sigma ^*(k\omega )= {T \over {2\nu }} \sum _{\tilde Q} G(\tilde
K+\tilde Q) {f_0^s}^2 \chi ^*(\tilde Q) \,. 
\end{equation}
The FP value $\chi ^*(\tilde Q)$ of the density-density response function is
determined by (\ref{chi}). \cite{ddrf} $\Sigma ^*(k\omega )$ is the
self-energy one would obtain in perturbation theory by dressing the particle
line with one density fluctuation propagator $\chi ^*(\tilde Q)$. It is
usually in this way that the quasi-particle properties are calculated in the
microscopic FLT. \cite{Stamp92} 

The quasi-particle life-time is obtained from the retarded part of the
self-energy: 
\begin{equation}
{1 \over \tau } \sim -{\rm Im}\, \Bigl \lbrace \Sigma ^*(k\omega ) \Bigl \vert 
_{i\omega \to \omega +i0^+} \Bigr \rbrace \,.
\end{equation}
For states close to the Fermi surface ($k,\omega \to 0$), this equation
yields the standard result for a two-dimensional Fermi liquid: $\tau ^{-1}
\sim T^2\ln T$. \cite{Hodges71,Bloom75,Stamp92} Since $\tau ^{-1}\ll T$ at
low temperature, the neglect of $\tau ^{-1}$ in the single-particle
propagator $G(\tilde K)$ during the renormalization is justified. 

From (\ref{RGEQ101},\ref{RGEQ11}), we obtain
\begin{equation}
d\ln (z)={\rm Re} \Biggl \lbrack 
{{\partial d\Sigma (k\omega )} \over {\partial i\omega }} \Biggl \vert
_{i\omega =k=0} \Biggr \rbrack \,,
\end{equation}
which yields 
\begin{eqnarray}
z^*&=&z_{\Lambda _0} \exp \Biggl \lbrace {\rm Re} \Biggl \lbrack 
{{\partial \Sigma ^*(k\omega )} \over {\partial i\omega }} \Biggl \vert
_{i\omega =k=0} \Biggr \rbrack \Biggr \rbrace 
\nonumber \\
&=& z_{\Lambda _0}
\exp \Biggl \lbrace -{T \over {2\nu }} \sum _{\tilde Q}
{\rm Re} \Biggl \lbrack
{{{f_0^s}^2\chi ^*(\tilde Q)} \over {(i\Omega -v_F^*\hat {\bf K}\cdot {\bf
Q})^2}} \Biggr \rbrack \Biggr \rbrace \,.  
\label{QPRF}
\end{eqnarray}
The preceding equation agrees with the result obtained from two-dimensional
bosonization \cite{Kopietz95} or Ward Identities. \cite{Castellani94}  
In the case of short-range interactions, the exponential factor in
(\ref{QPRF}) gives only a small correction to $z_{\Lambda _0}$ which can be
ignored for $\Lambda _0\ll K_F$.\cite{Kopietz95,Castellani94} This
ensures the existence of quasi-particles ($z^*>0$) for any non-vanishing
value of $z_{\Lambda _0}$.

\section{Conclusion}

We have shown in this paper how FLT results can be derived in a RG approach.
While it seems difficult to calculate physical quantities as a function of
the bare parameters of the low-energy effective action, it appears possible
(and quite natural) to relate them to the FP value of the $\Omega $-limit of
the two-particle vertex function which therefore determines the Landau
parameters. This result follows from the 
assumption that the ZS graph is the only singular graph in the limit $\tilde
Q \to 0$ and also the only one which is dominated by the integration of
low-energy states. These assumptions seem reasonable in cases where the
``quantum'' degrees of freedom ($\vert \epsilon \vert \gtrsim T$) do not
lead to any instability (such as superconductivity or charge/spin density
wave). As we pointed out, these two assumptions also underlie  
the standard diagrammatic derivation of FLT.

\section*{Acknowledgments}

I am indebted to C. Bourbonnais for many discussions which have strongly
contributed to my understanding of the RG approach to interacting
fermions. I thank G. Chitov for many stimulating discussions and a critical
reading of the manuscript. Useful discussions with L. Hubert, H. Schulz,
V. Yakovenko and D. Zanchi are also gratefully acknowledged.

\begin{references}
\bibitem{masse} The effective mass $m$ is not an independent parameter of the 
theory since it is related to the bare mass $m_0$ of the particles by 
$m/m_0=1+F^s_1/3$ as required by Galilean invariance.
\bibitem{Shankar94a} The importance of $n$-point ($n\geq 6$) vertices has been
pointed out by Shankar in a preliminary version of
Ref. \onlinecite{Shankar94}.
\bibitem{ddrf} The density-density response function is $\chi _{\rho \rho
}=\chi ^*$ and is RG invariant like any physical quantity. Here we improperly 
call $\chi (\tilde Q,t)$ the density-density response function. Notice
however that $\chi (\tilde Q,t)$ represents the contribution to $\chi _{\rho
\rho }$ of the degrees of freedom $\vert k\vert \geq \Lambda (t)$.
\end{references}
\end{document}